\documentclass[twocolumn,onecolappendix]{aastex63}

\usepackage{makecell}
\usepackage{xcolor}
\definecolor{dgreen}{RGB}{26,148,49}
\definecolor{xlinkcolor}{cmyk}{1,1,0,0}
\hypersetup{linkcolor=xlinkcolor,citecolor=xlinkcolor,urlcolor=xlinkcolor}

\usepackage{amsmath,amsthm,amssymb,amsfonts}
\usepackage{graphics}
\usepackage{listings}
\usepackage{enumitem}
\usepackage{url}
\usepackage{appendix}
\usepackage{hyperref}
\usepackage{mathtools}

\usepackage[super]{nth}

\newcommand{\N}{\mathbb{N}}
\newcommand{\Z}{\mathbb{Z}}

\newcommand{\zem}{\texttt{Z14}}

\newcommand{\bootes}{Bo{\"o}tes}

\begin{document}

\correspondingauthor{Richard M. Feder}
\email{rmfeder@berkeley.edu}

\author[0000-0002-9330-8738]{Richard M. Feder}
\affiliation{California Institute of Technology, 1200 E California Blvd, 91125, USA}
\affiliation{Berkeley Center for Cosmological Physics, University of California, Berkeley, CA 94720,
USA}
\affiliation{Lawrence Berkeley National Laboratory, Berkeley, California 94720, USA}

\author[0000-0002-5710-5212]{James J. Bock}
\affiliation{California Institute of Technology, 1200 E California Blvd, 91125, USA}

\author[0000-0002-5437-0504]{Yun-Ting Cheng}
\affiliation{California Institute of Technology, 1200 E California Blvd, 91125, USA}

\author[0000-0002-3892-0190]{Asantha Cooray}
\affiliation{Department of Physics and Astronomy, University of California, Irvine, CA 92697, U.S.A.}

\author{Phillip M. Korngut}
\affiliation{California Institute of Technology, 1200 E California Blvd, 91125, USA}

\author[0000-0002-5698-9634]{Shuji Matsuura}
\affiliation{School of Science and Technology, Kwansei Gakuin University, Sanda, Hyogo 669-1337, Japan}

\author[0000-0001-9368-3186]{Chi H. Nguyen}
\affiliation{California Institute of Technology, 1200 E California Blvd, 91125, USA}

\author[0000-0002-8405-9549]{Kohji Takimoto}
\affiliation{Department of Solar System Sciences, Institute of Space and Astronautical Science, Japan Aerospace Exploration Agency, 3-1-1 Yoshinodai, Chuo-ku, Sagamihara, Kanagawa 252-5210, Japan}

\author[0000-0001-8253-1451]{Michael Zemcov}
\affiliation{Center for Detectors, School of Physics and Astronomy, Rochester Institute of Technology, 1 Lomb Memorial Drive, Rochester, New York 14623, USA}
\affiliation{Jet Propulsion Laboratory, California Institute of Technology, 4800 Oak Grove Drive, Pasadena, CA 91109, USA}

\author{CIBER collaboration}

\title{\emph{CIBER} \nth{4} flight fluctuation analysis: Pseudo-power spectrum formalism, improved source masking and validation on mocks}
\begin{abstract}
    Precise, unbiased measurements of extragalactic background anisotropies require careful treatment of systematic effects in fluctuation-based, broad-band intensity mapping measurements. In this paper we detail improvements in methodology for the Cosmic Infrared Background ExpeRiment (\emph{CIBER}), concentrating on flat field errors and source masking errors. In order to bypass the use of field differences, which mitigate flat field errors but reduce sensitivity, we characterize and correct for the flat field on pseudo-power spectra, which includes both additive and multiplicative biases. To more effectively mask point sources at 1.1 $\mu$m and 1.8 $\mu$m, we develop a technique for predicting masking catalogs that utilizes optical and NIR photometry through random forest regression. This allows us to mask over two Vega magnitudes deeper than the completeness limits of 2MASS alone, with errors in the shot noise power remaining below $<10\%$ at all masking depths considered. Through detailed simulations of \emph{CIBER} observations, we validate our formalism and demonstrate unbiased recovery of the sky fluctuations on realistic mocks. We demonstrate that residual flat field errors comprise $<20\%$ of the final \emph{CIBER} power spectrum uncertainty with this methodology.
\end{abstract}

\keywords{cosmology: Diffuse radiation –- Near infrared astronomy -- Large-scale structure of universe -- Galaxy evolution -- Cosmic background radiation}


\section{Introduction}
\label{S:introduction}
The extragalactic background light (EBL) is the integrated light from all sources outside of the Milky Way, emitted over cosmic history. The spectral and spatial characteristics of the EBL promise a wealth of information on the astrophysical processes that drive cosmic light production. However, at optical and near-infrared (NIR) wavelengths, both the measurement and interpretation of the EBL have not converged, with disagreements between various methods \citep{Cooray2016, Carleton2022, Hill2018, Matsuura2017, Akshaya2019, HESS2017, Fermi2019, Kashlinsky2012}. 

Fluctuation-based measurements of the EBL bypass the conventional challenge of absolute photometric measurements, namely degeneracy with zodiacal light (ZL), taking advantage of its smoothness on large scales as measured at infrared and mid-IR wavelengths \citep{arendt16, abraham_97}. However, fluctuation measurements are sensitive to other systematics which need careful treatment. In our previous work (\citealt{zemcov14}, hereafter \zem) we measured fluctuations at 1.1 $\mu$m and 1.6 $\mu$m using imaging data from the second and third flights of \emph{CIBER-1} (the first generation of \emph{CIBER}), revealing fluctuations on angular scales $\theta > 5^{\prime}$ with an amplitude exceeding that expected from integrated galactic light (IGL), although without accounting for non-linear clustering \citep{cheng22}.

In this work we improve upon the methodology in \zem\ by focusing on two leading effects. The first involves corrections for the relative per-pixel gain within each imager, commonly known as the flat field gain (denoted FF). Contrary to \zem, which relied on field differences to mitigate FF errors, we directly estimate the FF gain from the science field observations with a stacking estimator and develop a pseudo-power spectrum formalism that quantifies and corrects for errors in the FF estimator. The second effect involves masking bright stars and galaxies, which is required to reduce Poisson fluctuations from bright sources. Through ancillary optical and infrared photometry, we mask substantially deeper than by using 2MASS JHK$_s$ photometry alone while minimizing the fraction of masked pixels.

Monte Carlo simulations play an important role in our power spectrum pipeline. Accurate simulations of noise present in the \emph{CIBER} maps (combined with realistic masking) allow us to estimate statistical errors and to correct for various noise biases on the power spectrum. By performing the same data processing on synthetic mocks as used for the observed data, we are able to validate our FF formalism and assess any remaining biases in the power spectrum pipeline. Our mock recovery tests on a large ensemble of synthetic \emph{CIBER} observations allow us to estimate uncertainties and covariances in a comprehensive fashion and enable an assessment of field-to-field consistency in the observed \emph{CIBER} data.

The paper is organized as follows. In \S 2 we introduce \emph{CIBER} and describe the construction of synthetic mocks in \S 3, which include known astrophysical components, realizations of \emph{CIBER} read and photon noise along with other observational effects. In \S 4 we describe the standard steps of the pseudo-power spectrum pipeline while in \S 5 we present the extended formalism that includes flat field errors. We then introduce a novel source masking procedure in \S 6 which includes several catalog-level validation tests. In \S 7 we apply our improved power spectrum pipeline to the mocks from \S 3, validating our ability to recover unbiased estimates of sky fluctuations and quantifying the impact of flat field errors on our power spectrum sensitivity. Lastly,  we conclude in \S 8 and discuss avenues for future development.

Throughout this work we assume a flat $\Lambda$CDM cosmology with $n_s=0.97$, $\sigma_8=0.82$, $\Omega_m=0.26$, $\Omega_b=0.049$, $\Omega_{\Lambda}=0.69$ and $h=0.68$, consistent with measurements from \emph{Planck} \citep{planck16}. All fluxes are quoted in the Vega magnitude system unless otherwise specified. 
\section{Cosmic Infrared Background Experiment (CIBER)}
\label{S:ciber}

\emph{CIBER}\footnote{\url{https://ciberrocket.github.io/}} is a rocket-borne instrument \citep{zemcov13} designed to characterize the NIR EBL through measurements of its spatial fluctuations and electromagnetic spectrum \citep{bock13, ciber_lrs}. In this work we focus on measurements using the \emph{CIBER} imagers, simultaneously observing a 2 $\times$ 2 deg$^2$ field of view in two broad bands centered at 1.1 $\mu$m and 1.8 $\mu$m with 7\arcsec\ pixels with wide-field refracting optics.

During the \emph{CIBER}-1 flight integration campaigns, laboratory FF measurements were conducted using an integrating sphere for uniform illumination with a solar-type spectrum. However, these measurements were inconsistent with the FF estimates derived from flight exposures. We attribute the difference to systematic errors in the lab measurement (near field of the optics, non-uniformity of the sphere, residual spectral mismatch, etc). \zem\ therefore analyzed field differences, and used the difference in the FFs between laboratory measurements and flight data to estimate the residual FF uncertainty. 

\emph{CIBER}-1 was flown four times in total, the first three from White Sands Missile Range in New Mexico, and the final non-recovered flight from NASA Wallops Flight Facility in Virginia. Unlike during the first three flights, the payload during the fourth flight achieved an altitude of 550 km (compared to $\sim 330$ km), resulting in a longer total exposure time and lower levels of airglow contamination. Crucially, the higher number of science exposures from the fourth flight (five science fields compared to 2-3 for previous flights) enables the use of an improved in-flight FF stacking estimator for which per-pixel errors are sufficiently small, a condition we formalize in this work. Relaxing the requirement of field differences reduces the fraction of masked pixels in each map, which allows us mask more aggressively on individual fields, given access to sufficiently deep external catalogs.

\section{Simulations of CIBER Observations}
\label{Sec:mocks}

The synthetic observations described in this section serve to validate the power spectrum estimation pipeline and to estimate covariances which are then used against real datasets in Paper II. We generate mock sky realizations with a combination of point-source and diffuse clustering components. This includes random source realizations of galaxies and stars, diffuse galactic light (DGL), ZL, and EBL clustering fluctuations to match the observed auto-power spectrum measurements of \zem. Our synthetic observations match the characteristics of \emph{CIBER}-1's fourth flight imaging dataset, including read and photon noise (including one read noise dominated science exposure with less than half the exposure time of the other four fields), pixel scale and point spread function (PSF). While the mock IGL and EBL clustering are taken to be statistically similar across the five \emph{CIBER} fields, both the ZL and ISL vary due to the range of ecliptic and galactic latitudes spanned by the fields. 

\subsection{CIB galaxies}
To generate IGL mocks we combine log-normal realizations of the matter density field with Poisson draws on empirical, redshift-dependent luminosity functions from \cite{helgason}. The log-normal mock technique has been developed for fast generation of galaxy catalogs and density fields \citep{galaxyclusgen, lognormal}. For a desired power spectrum $C_{\ell}$, we first generate a field $G(x)$ from the log-normal power spectrum, denoted $C_{\ell}^G$, and exponentiate the field, i.e., $\delta(x) \sim \exp\left[G(x)\right]$. To compute $C_{\ell}^G$, we compute the angular two-point correlation function for $C_{\ell}$ using the Hankel transform
\begin{equation}
    w(\theta) = \int \frac{\ell d\ell}{2\pi} C_{\ell}J_0(\ell \theta).
\end{equation}
In the above equation, $J_0$ is a Bessel function of the zeroth kind and we have invoked the flat-sky approximation, which is valid on the scales considered. We compute the angular power spectrum $C_{\ell}$ from the projected non-linear, redshift-dependent 3D matter power spectrum using the Python version of \texttt{CAMB}. Next, we transform the angular two-point correlation function into the log-normal correlation function
\begin{equation}
    w^G(\theta) = \log(1+w(\theta)).
\end{equation}
The log-normal correlation function is then converted back to an angular power spectrum using the inverse Hankel transform:
\begin{equation}
    C_{\ell}^G = 2\pi \int \theta d\theta w^G(\theta)J_0(\ell \theta).
\end{equation}
The log-normal power spectrum defines the diagonal component of the covariance matrix $\pmb{C}_{\ell}^G$. Finally to generate the field $G(x)$, we draw a Gaussian realization from $\mathcal{N}(0, \pmb{C}_{\ell}^G)$ and compute its inverse Fourier transform, discarding the imaginary component. This is then exponentiated to obtain the nonlinear density field $\delta(x)$. To reliably simulate fluctuation modes on the scale of the \emph{CIBER} fields, we generate realizations of $\delta(x)$ over a larger $4\times4$ deg$^2$ field after which we extract the central $2\times2$ deg$^2$ regions. 

We then generate projected galaxy counts as Poisson realizations of each underlying density field, in which the mean number of galaxies per cell is set by the integrated number counts, i.e.,
\begin{equation}
    N_{tot,i} = \int_{z_{i}}^{z_{i+1}} dz \frac{dN}{dz}.
\end{equation}
We simulate independent IGL realizations in eight equally-spaced redshift bins between $0\leq z \leq 2$. Note that these log-normal mocks do not include the effect of galaxy biasing which enhances the large-scale power from two-halo clustering; however, as the total observed \emph{CIBER} fluctuations on large scales exceeds that from IGL by over an order of magnitude (with and without galaxy biasing), we do not incorporate a detailed biasing scheme in the IGL mocks. We refer the reader to \cite{cheng22} for the effects of galaxy bias and non-linear IGL clustering as predicted using MICECAT simulations.

We use the semi-empirical model from \cite{helgason} to produce realizations of the IGL. The Helgason model constructs galaxy luminosity functions (LFs) assuming a Press-Schechter functional form fit to a complilation of observed LFs from existing measurements. Within each redshift bin we assign galaxy redshifts from the normalized $dN/dz$ distribution, conditioned on the LF at each bin center:
\begin{equation}
    \frac{dN}{dz} = \int \Phi(M|\hat{z}_i) dM.
\end{equation}
Using these LFs we then draw apparent magnitudes for each source down to $m_{AB}=28$, corresponding to different absolute magnitudes $M_{abs}^{min}$ within each redshift bin.


The fits from \cite{helgason} have uncertainties related to the faint end slopes of the LFs, from which they delineate ``High Faint End" (HFE) and ``Low Faint End" (LFE) model predictions. To assess the impact of these uncertainties on the predicted Poisson noise level we generate three separate sets of CIB mocks by varying $\alpha_0$, the Schechter parameter that normalizes the faint end slope, i.e.,
\begin{align}
    \phi(M)dM &\propto \phi^*\left(10^{0.4(M^*-M)}\right)^{\alpha(z)+1},
\end{align}
where $\alpha(z) = \alpha_0(z/z_0)^r$ \citep{helgason}. We use $\alpha_0=-1.0$ (default), $\alpha_0=-0.8$ (LFE) and $\alpha_0=-1.2$ (HFE) and compute the corresponding power spectra with a range of masking depths. On small scales, the HFE models predict 15-25\% higher power between $J=17.0$ and $J=18.5$ than the default model, while the LFE models predict $5-10\%$ less power. The models differ most on large scales, with HFE a factor of $1.5-2$ larger than the default model for $\ell < 1000$. We use the fiducial model prediction for the results in this work. This IGL model has the limitation that only single bands can be simulated, i.e., we cannot simulate or make predictions for cross-spectra across different wavelengths. 

\subsection{Zodiacal light}
ZL refers to light reflected off interplanetary dust grains (IPD) within our solar system. Beyond Earth's atmosphere, ZL is the leading contribution to the intensity monopole at NIR wavelengths. \cite{kelsall} constrains the ZL contribution from DIRBE observations across $1.25-240$ $\mu$m with a precision of $\sim 1\%$, under the assumption that the IPD should be the only time-varying component on the celestial sphere. The ZL is measured to be spatially smooth on degree scales by DIRBE and smooth on scales $\theta \lesssim 200\arcsec$ by \emph{Spitzer} \citep{arendt16}. We use a modified Kelsall model that accounts for the solar spectrum and ZL reddening to predict intensity over each \emph{CIBER} bandpass \citep{Crill2020}. The ZL intensity varies by a factor of 2.3 across the five \emph{CIBER} fields, spanning ecliptic latitudes $11^{\circ} \leq \beta \leq 73^{\circ}$. In addition to the monopole, we inject random ZL gradients into the mocks with amplitudes derived from K98. However we give the gradients random directions, to capture the effect of the high-pass image filtering used in the pipeline.


\subsection{Integrated Stellar Light}
\label{sec:trilegal}

The raw \emph{CIBER} fluctuation power is dominated by Poisson fluctuations from bright stars within our galaxy. We use the TRILEGAL model \citep{trilegal} to simulate realistic distributions of stars within each \emph{CIBER} field. These are also used in \cite{chengihl} and are useful for testing the efficacy of astronomical masking in the maps and also for estimating the contribution of integrated star light below the masking threshold. On degree scales and at high galactic latitudes, the angular distribution of the ISL is well approximated as uniform \citep{zemcov14}. However, the ISL amplitude varies across the five \emph{CIBER} fields, with SWIRE having $\sim 50\%$ higher stellar density compared to the mean of the fields.

\subsection{Diffuse galactic light and EBL clustering components}
We include an additional clustering signal from Gaussian realizations with power spectrum of the form $C_{\ell} = A\ell^{-3}$, where $A$ is chosen to match the observed \emph{CIBER} power spectrum in \zem. This component has the same spatial index as expected for DGL, an important foreground for intensity mapping at NIR wavelengths. However, \zem\ showed that the DGL component was small compared to the total signal. We simply model the total \zem\ signal to properly estimate biases in power spectrum recovery, sample variance and covariance. The simulated Gaussian realizations neglect potential correlations between the EBL signal and simulated CIB galaxies.

\subsection{Synthetic CIBER observations}

We model the extended \emph{CIBER} fourth flight PSF to populate point sources in each mock. The measured PSF is a combination of the instrumental PSF and the pointing jitter and drift over each exposure. \cite{chengihl} estimated field-dependent PSFs by stacking \emph{CIBER} images on the positions of 2MASS sources and modeling the profiles with a $\beta$ model of the form
\begin{equation}
    \textrm{PSF}(r) = \left(1 + \left(\frac{r}{r_c}\right)^2\right)^{-3\beta/2}.
\end{equation} 
We evaluate the \emph{CIBER} PSF at 100 sub-pixel positions and use these templates to inject sources into the maps at native resolution. Capturing the sub-pixel PSF is important for \emph{CIBER}, which is designed to have an undersampled beam (the instrument PSF has a FWHM$\sim 9\arcsec$, slightly more than one \emph{CIBER} pixel). Due to the relatively coarse \emph{CIBER} angular resolution, we populate galaxies into the mocks as point sources and do not explicitly simulate the small number of extended extragalactic objects as these are identified and masked appropriately from catalogs. We do not include effects of PSF variation across the focal plane; however, this is an important effect in more precise measurements of small-scale ($\ell > 50000$) fluctuations.

Lastly, we add noise realizations to each sky realization. Our noise model is constructed from two components. The first is read noise from the detector and readout electronics, which we estimate from dark exposures obtained just prior to launch. Specifically, we generate Gaussian realizations with random phases from an underlying two-dimensional power spectrum describing each integration, which is necessary to capture the anisotropic structure of the read noise. The second is photon noise due to the Poisson statistics of sky signal incident on the \emph{CIBER} imagers. We use the model of \cite{garnett_forrest} to calculate the Poisson variance map for each sky realization:
\begin{equation}
    \sigma_{\gamma}^2 = \frac{6}{5}\frac{i_{phot}}{T_{int}}\left(\frac{N^2+1}{N^2-1}\right),
\end{equation}
where $i_{phot}$ is the photocurrent, $T_{int}$ is the integration time and $N$ is the number of frames in each integration.

\begin{figure*}
    \centering
    \includegraphics[width=0.48\linewidth]{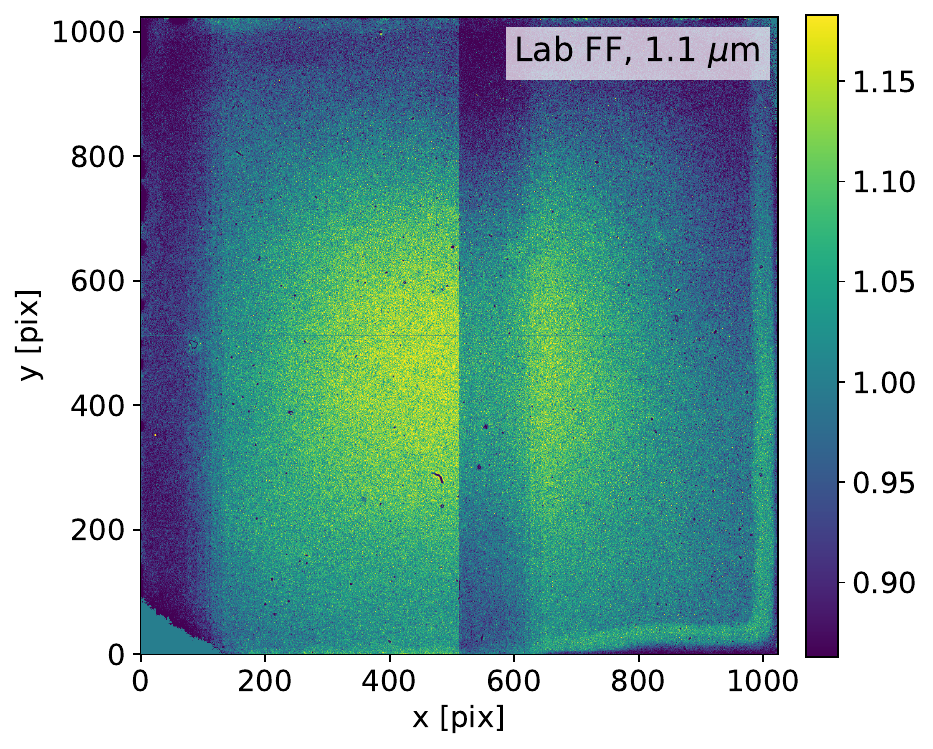}
    \includegraphics[width=0.48\linewidth]{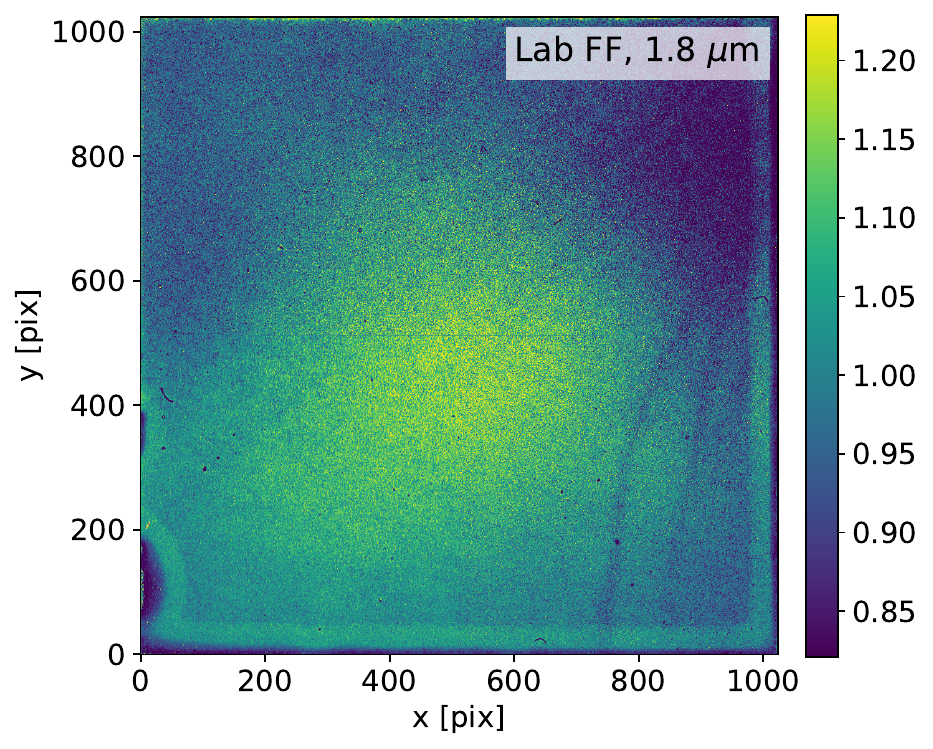}
    \caption{Laboratory FF measurements taken during the third \emph{CIBER}-1 flight campaign. Structure in the \emph{CIBER} FFs comes from a combination of optical and electrical effects. We use the laboratory data to inject a realistic FF into our mocks, which is then estimated and corrected for in our power spectrum recovery tests (see \S \ref{sec:mock_ps_test}).}
    \label{fig:lab_ff}    
\end{figure*}

While our formalism for recovering sky fluctuations is insensitive to the exact shape of the true FF, we use the laboratory FF estimates from \emph{CIBER}-1's third flight campaign to realistically modulate each sky signal (and its photon noise), after which we add read noise which does not depend on the FF. We show the laboratory flats for the 1.1 $\mu$m and 1.8 $\mu$m detectors in Fig. \ref{fig:lab_ff}.

\begin{figure*}
    \centering
    \includegraphics[width=0.95\linewidth]{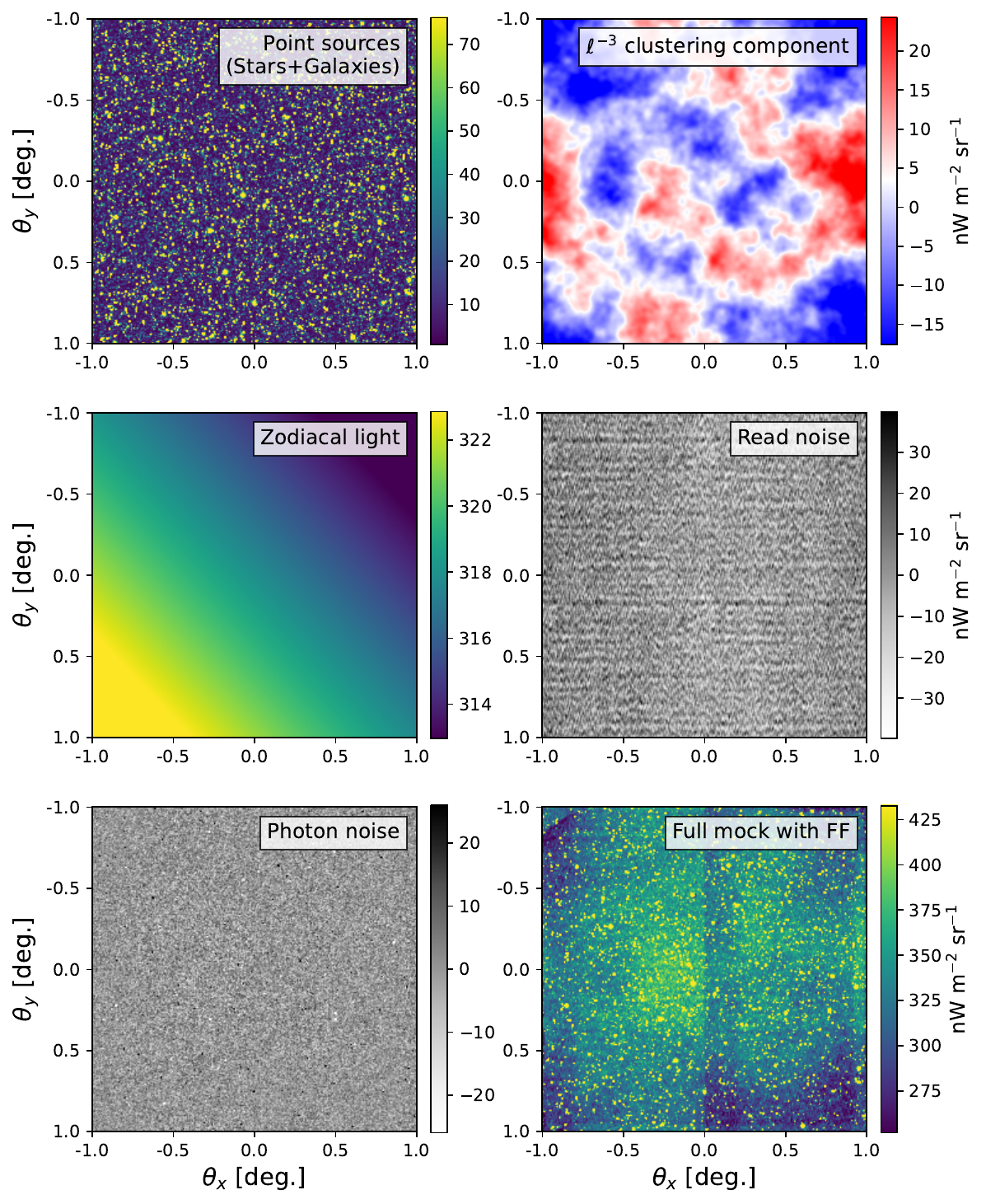}
    \caption{Different astrophysical signal and noise components that compose the mock \emph{CIBER} observations used in this work, shown for 1.1 $\mu$m in sky units (nW m$^{-2}$ sr$^{-1}$). We use these mocks to simulate power spectrum recovery on realistic synthetic data.}
    \label{fig:multipanel_mock}
\end{figure*}

We show the signal and noise components that go into each \emph{CIBER} mock observation in Fig. \ref{fig:multipanel_mock} for 1.1 $\mu$m. 

\section{Power spectrum formalism}
\label{Sec:ps_estimation}

In this section we describe the steps to relate an underlying sky intensity power spectrum to the binned, observed pseudo-power spectrum. We first present the calculation assuming no FF errors, and detail the modifications given our FF stacking estimator in the subsequent section.

The observed power spectrum can be related to the true sky power spectrum as
\begin{equation}
    C_{\ell}^{obs} = \sum_{\ell'}M_{\ell\ell^{\prime}}(B_{\ell'}^2 C_{\ell'}^{sky} + N_{\ell'}).
    \label{eq:image_form}
\end{equation}
The sky signal with power spectrum $C_{\ell}^{sky}$ (shorthand for all astrophysical components) is passed through the imaging process, which smoothes the signal on the scale of the beam. The masking, FF correction and filtering we employ couple large- and small-scale modes in the maps and enhance and/or suppress fluctuations. We capture these effects using a linear mode mixing matrix $M_{\ell\ell'}$. In \eqref{eq:image_form} $B_{\ell}$ is the beam transfer function, which is assumed to be diagonal in Fourier space, and $N_{\ell}$ is the noise bias. To recover $C_{\ell}^{sky}$ we correct for the noise bias, apply a mode coupling correction $M_{\ell \ell^{\prime}}^{-1}$ and then correct for the beam transfer function,
\begin{equation}
    \hat{C}_{\ell}^{sky} = B_{\ell}^{-2}\sum_{\ell'}M_{\ell\ell'}^{-1}(\hat{C}_{\ell'}^{obs} - \hat{N}_{\ell'}).
    \label{eq:invert_cl}
\end{equation}

While we use $\ell, \ell'$ for presentation, in practice we implicitly apply a binning operator $P_{b\ell}$ to average modes into bandpowers $b$, i.e.,
\begin{equation}
    \hat{C}_{b} = \sum_{\ell \in \mathcal{B}} P_{b\ell} \hat{C}_{\ell},
\end{equation}
where $\mathcal{B}$ denotes the set of modes $\ell'$ satisfying $\ell_b^{min} \leq \ell' < \ell_b^{max}$. For uniform weighting the binning operator is
\begin{equation}
    P_{b\ell} = \begin{cases}
        \frac{1}{\ell_b^{max}-\ell_b^{min}},\quad \ell \in \mathcal{B} \\
        0,\quad \ell \notin \mathcal{B}
    \end{cases}.
\end{equation}

\subsection{Noise bias subtraction}
\label{sec:noise_bias_standard}
A positive noise bias $N_{\ell}$ arises in the auto-power spectrum due to instrument noise fluctuations, which also mix with the image mask:
\begin{equation}
    N_{\ell} = \sum_{\ell'}M_{\ell\ell'}(N_{\ell'}^{read}+N_{\ell'}^{\gamma}).
\end{equation}
The per-pixel photon noise $N_{\ell'}^{\gamma}$ depends on the beam-convolved sky maps. For each field we estimate $N_{\ell}$ using an ensemble of 500 independent noise realizations combined with the respective image mask. This is typically done through draws of read and photon noise alone, however as discussed in \S \ref{sec:ff_noise_bias} there are additional contributions to the noise bias from instrument noise-driven errors in the stacked FFs.

\subsection{Fourier weighting}

Detector read noise power in the maps can be mitigated by recognizing that certain 2D Fourier modes contribute significantly more noise power than others. By calculating the per-mode variance of our Monte Carlo noise realizations from the previous sub-section, we derive inverse variance weights to the two-dimensional power spectrum which we apply before computing azimuthally-averaged bandpowers,
\begin{equation}
    \langle C_{\ell} \rangle = \frac{\sum_{(\ell_x, \ell_y)} w(\ell_x, \ell_y)C(\ell_x, \ell_y) }{\sum_{(\ell_x, \ell_y)} w(\ell_x, \ell_y)},
\end{equation}
where $C(\ell_x,\ell_y)$ is the two-dimensional observed power spectrum. There is a trade off between down-weighting noisy Fourier modes and reducing the effective sample size of modes contributing to each bandpower. In addition, the mask convolution tends to spread concentrated power across adjacent modes.  Despite these caveats, we find that 2D Fourier weighting is effective at mitigating variance from \emph{CIBER} read noise fluctuations, which are highly anisotropic in the Fourier plane.





\subsection{Beam correction}

The \emph{CIBER} PSF smoothes the observed sky signal, resulting in a roll-off in power on small scales. Using the best-fit beam model for each field, we generate a $10\times 10$ grid of sub-pixel centered PSFs, which are downsampled to the \emph{CIBER} native pixel resolution. We then compute $B_{\ell}^2$ as the mean power spectrum of the 100 sub-pixel PSFs.

\subsection{Mode coupling correction}

The application of instrument and astronomical masks on the \emph{CIBER} maps means that Fourier modes on the underlying sky will mix with one another. Following the MASTER formalism \citep{master}, we estimate each $N_{bp} \times N_{bp}$ bandpower mode coupling matrix by applying the target field mask to Monte Carlo tone realizations with random phase through the target mask, and compute the corresponding pseudo-power spectra. We do not apply any apodization to the images. In this work we fix the number of logarithmically-spaced bandpowers to $N_{bp} = 25$. For each bandpower we compute $N_{sim}=500$ phase realizations, where $N_{sim}$ is chosen to be large enough such that statistical errors on $\hat{M}_{\ell \ell^{\prime}}$ are negligible. After taking the expectation across realizations, we correct the observed, noise-debiased power spectrum by applying the inverse mode coupling matrix $M_{\ell \ell^{\prime}}^{-1}$.


\subsection{Image filtering}

The \emph{CIBER} maps have large-scale variations which need to be filtered out before computing power spectra. The first are array-scale gradients which come from ZL and other foreground components. The second involves a quadrant-specific detector effect we identified in both laboratory and flight data. In some quadrants, we observe a form of two-state noise, in which the ADU levels in all pixels fluctuate coherently across consecutive readout frames. This leads to a variation in the resulting slope fits across quadrants. We correct for these effects by fitting a linear combination of per-quadrant offsets and a gradient across the full array,
\begin{equation}
    G(x,y) = Ax + By + \sum_{k=1}^4 Q_k O_k(x,y),
\end{equation}
where $A$ and $B$ are the gradient parameters, $Q_k$ is the offset parameter for quadrant $k$ and $O_k(x,y)$ is a 2D step function equal to one for pixels in quadrant $k$ and zero otherwise. 

In this work we also consider a more aggressive image filter, in which we fit a set of templates representing individual low-$\ell$ Fourier modes to the maps. This model basis has been used in previous work separating diffuse and pointlike signals in \emph{Herschel}-SPIRE images \citep{pcatde}. Using a set of ``global" Fourier templates differs from polynomial filters that depend on local derivatives of the image. In this work we argue that the Fourier component model more effectively separates large-angle fluctuations in the observed CIBER maps. 

The 2D truncated Fourier series model $B(x,y)$ can be written as
\begin{equation}
    B(x,y) = \sum_{n_x=1}^{\textrm{N}_{\textrm{FC}}}\sum_{n_y=1}^{\textrm{N}_{\textrm{FC}}}\pmb{\beta}_{n_x n_y}\cdot \pmb{\mathcal{F}}(x,y)^{n_x n_y}.
    \label{eq:totalbkg}
\end{equation}
In this equation, $\textrm{N}_{FC}$ refers to the order of the Fourier series and $\pmb{\mathcal{F}}(x,y)^{n_x n_y}$ is a vector of Fourier components corresponding to wavevector $(k_x, k_y) = (\textrm{W}/n_x,\textrm{H}/n_y)$, where W and H denote the image dimensions, evaluated at pixel $(x,y)$:
\begin{equation}
    \pmb{\mathcal{F}}(x,y)^{n_x n_y} = \left(\begin{array}{c} \sin\left(\frac{n_x\pi x}{\textrm{W}}\right)\sin\left(\frac{n_y\pi y}{\textrm{H}}\right) \\ 
    \sin\left(\frac{n_x\pi x}{\textrm{W}}\right)\cos\left(\frac{n_y\pi y}{\textrm{H}}\right)
    \\
    \cos\left(\frac{n_x\pi x}{\textrm{W}}\right)\sin\left(\frac{n_y\pi y}{\textrm{H}}\right)
    \\
    \cos\left(\frac{n_x\pi x}{\textrm{W}}\right)\cos\left(\frac{n_y\pi y}{\textrm{H}}\right)
    \end{array}\right).
    \label{eq:fcomp}
\end{equation}
Each $\pmb{\beta}_{n_x n_y}$ contains the coefficients for the four sinusoidal components corresponding to $(n_x, n_y)$. We arrive at a model with $\textrm{N}_{FC}=2$, corresponding to sixteen templates, which removes the largest scale power but does not fully suppress power on sub-degree to few arcminute scales. 


We calculate the best-fit parameters $\theta$ from the least-squares solution to unmasked pixels $\vec{K}$ with the Moore-Penrose pseudo-inverse,
\begin{equation}
    \vec{\theta} = (X^T X)^{-1}X^T \vec{K}.
    \label{eq:quadoff_grad}
\end{equation}

\begin{figure}
    \centering
    \includegraphics[width=\linewidth]{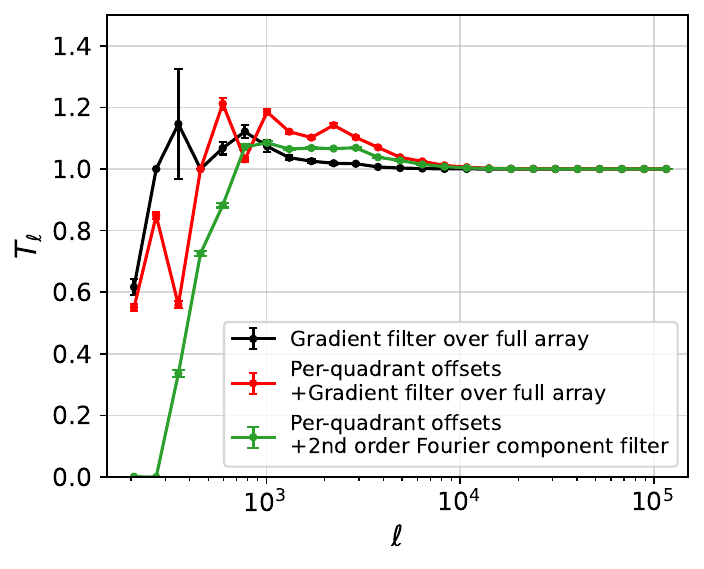}
    \caption{Comparison of 1D filter transfer functions, each estimated using the ratio of filtered and original power spectra of 1000 Gaussian signal realizations. Errorbars indicate the dispersion across realizations.}
    \label{fig:t_ell}
\end{figure}

For illustrative purposes in Fig. \ref{fig:t_ell} we plot the 1D transfer functions $T_{\ell}$ of three filtering configurations. These are derived from 1000 Monte Carlo signal realizations that include both large- and small-scale power, for which we calculate the ratio of power spectra before and after filtering. To check for potential dependence on the input clustering, we test $T_{\ell}$ with three separate spatial indices ($\gamma \in \lbrace -2, -2.5, -3\rbrace$ where $C_{\ell} \propto \ell^{\gamma}$ on large scales) and find estimates of $T_{\ell}$ for each case that are consistent within statistical uncertainties. In all cases there is a suppression of power in the lowest bandpowers as well as an increase in power for intermediate bandpowers. Including per-quadrant offsets to the gradient fitting step leads to a form of ringing that can be seen in the shape of $T_{\ell}$ across adjacent bandpowers. The effect is reduced, however, when the gradient filter is replaced with the \nth{2}-order Fourier component model filter. The Fourier component filter nearly completely suppresses the lowest two bandpowers and more aggressively removes power compared to the other two filters on scales $\ell \lesssim 800$.    

Although it is a common ansatz to treat the filtering transfer function as a 1D quantity $T_{\ell}$ that is applied after mask de-convolution, we find non-negligible mode coupling introduced by image filtering. For this reason, we ultimately incorporate the filtering operation into the mixing matrix formalism, described in \S \ref{sec:mode_couple_ff}.

\section{Power spectrum biases from the flat field stacking estimator}
\label{Sec:ff_bias_formalism}

\subsection{FF estimation and image filtering}


Throughout this section we express the true sky signal, denoted $I_{sky}$, in terms of a mean intensity $\overline{I}_{sky}$ and a general fluctuation component $S$,
\begin{equation}
    I_{sky} = \overline{I}_{sky} + S.
\end{equation}

The FF gain across each detector array is defined as the relative response to a uniform illuminating surface, which in practice is determined by detector effects (e.g., variations in per-pixel quantum efficiency), as well as the optical/mechanical configuration of the instrument, which can introduce effects such as vignetting. Described by a scalar field $FF(x,y)$, the flat field modulates the incident sky signal
\begin{equation}
    I_{sky}(x,y) \to FF(x,y)I_{sky}(x,y).
\end{equation}
In the absence of a well-determined FF, \zem\ used field differences to mitigate errors in the FF at leading order, following the fact that FF errors primarily couple to the mean intensity of each map:
\begin{align}
    \delta I_{A-B} &= \delta[\hat{FF}](I_A^{obs}-I_B^{obs})\\
    &\approx \delta[\hat{FF}](\overline{I}_A - \overline{I}_B).
\end{align}

Field differences also make power spectrum de-convolution more difficult due to higher masking fractions (the effective mask is the union of individual field masks), degrading statistical sensitivity and limiting the achievable masking depth for point sources. This penalty is especially pronounced for some measurements, e.g., cross-correlations with \emph{Spitzer} data in single fields \citep{zemcov14}. 

The alternative approach we pursue in this work is to estimate the FF from exposures taken during flight, stacking the FF images derived from each field. We utilize the fact that the mean background of each field (which is dominated by ZL) acts as an approximate uniform illuminator. For field $i$, the FF image is defined as
\begin{equation}
    \hat{FF}_i(x,y) = \frac{I_i^{obs}(x,y)}{\overline{I}_i^{obs}}.
\end{equation}
While the astrophysical signals will vary across images, the FF responsivity does not, and so in the limit of many independent exposures this estimator should converge to the true FF,
\begin{equation}
    \lim_{N_f\to \infty} \langle \hat{FF}(x,y)\rangle = FF_{true}(x,y).
\end{equation}
We model the FF error in unmasked pixels from a single field as
\begin{equation}
    \delta \hat{FF}_i(x,y) = \frac{FF(x,y)(S_i + \epsilon_{\gamma,i}) + \epsilon_{read,i}}{\overline{I}_i^{obs}}.
    \label{eq:ff_error}
\end{equation}
where $S_i$ is the underlying sky signal, $\epsilon_{\gamma}$ is the photon noise and $\epsilon_{read}$ is the read noise. 

The FF estimate for each target field $j$ comes from stacking the $N_{f}=4$ other science fields (``off-fields") in a round-robin approach,  
\begin{equation}
    \hat{FF}_j(x,y) = \frac{\sum_{i=1}^{N_f}w_i \hat{FF}_i}{\sum_i w_i}.
\end{equation}
We apply field weights $w_i$ that account for both photon and read noise to construct a minimum variance per-pixel FF estimate. The loss of pixels due to masking in the FF images means each field does not contribute an FF estimate for all pixels. There is a small ($\sim 1\%$) fraction of pixels that by chance have zero unmasked $FF_j$ estimates, in which case we mask these pixels in the $j$th science field. To avoid non-linear effects sourced from large FF errors, we additionally mask pixels with FF estimates that deviate by $>3\sigma$ from the mean local FF estimate, which affects $1-2\%$ of pixels depending on the field.

By performing the FF stacking on mocks, we find that the per-pixel error RMS is $\sigma(\delta FF) \sim 4-5\%$ for both imagers, with $>94\%/97\%$ of pixels having $|\delta FF| < 0.1$ for 1.1 $\mu$m and 1.8 $\mu$m respectively. These errors are largely driven by instrumental noise. Due to our round robin stacking approach, each field has a different FF error distribution. 

\subsection{Power spectrum bias}
We proceed with this estimator by expressing $\hat{FF}^j$ in terms of the true FF and its error, such that 
\begin{equation}
   \frac{I_j^{obs}}{\hat{FF}^j} = \frac{FF_{true}\left[I_j^{sky}+\epsilon_{\gamma,j}\right] + \epsilon_{read,j}}{FF_{true}\left[1 + \frac{\delta [\hat{FF}^j]}{FF_{true}}\right]}.
   \label{eq:Iobs_psbias}
\end{equation}
Taking the limit where $\delta[\hat{FF}^j]/FF_{true} \ll 1$, we Taylor expand Eq. \eqref{eq:Iobs_psbias} and in App. \ref{sec:ff_formalism} arrive at the following expression for the corrected map,
\begin{multline}
    \frac{I_j^{obs}}{\hat{FF}^j} \approx \overline{I}_j^{sky} + S_j + \epsilon_{\gamma,j} + \frac{\epsilon_{read,j}}{\hat{FF}^j} \\ - \frac{\delta[\hat{FF}^j]}{\hat{FF}^j}(\overline{I}_j^{sky}+S_j+\epsilon_{\gamma,j}).
    \label{eq:ff_corr_Iobs}    
\end{multline}
The FF error terms in the second line above introduce additional biases on the sky power spectrum. The FF error coupled with the mean sky brightness of the target field, $\delta[\hat{FF}^j] \overline{I}_j^{sky}$, sources the majority of the fluctuation bias. While we simulate and correct for the point source contributions to $\delta[\hat{FF}^j](S_j + \epsilon_{\gamma, j})$, these comprise $<1\%$ of the total noise bias in each field and have a negligible impact at \emph{CIBER} sensitivity. 

From Eqs. \eqref{eq:ff_error} and \eqref{eq:ff_corr_Iobs} we see that the power spectrum FF bias depends on both instrument noise and sky fluctuations. While the instrument noise contribution yields an additive bias, the sky fluctuations act as a multiplicative bias. Another conclusion is that the amplitude of the PS bias depends on the relative mean sky brightnesses across fields, such that fields with higher ZL have larger biases.

\subsubsection{Modified noise bias}
\label{sec:ff_noise_bias}
FF errors sourced by instrument noise lead to an additional noise bias contribution, which we denote $N_{\ell}^{\delta FF}$,
\begin{equation}
    N_{\ell} = \sum_{\ell'}M_{\ell\ell'}(N_{\ell'}^{read}+N_{\ell'}^{\gamma} + N_{\ell'}^{\delta FF}).
\end{equation}
To include $N_{\ell^{\prime}}^{\delta FF}$ within the Monte Carlo procedure in \S \ref{sec:noise_bias_standard}, we add mean sky levels to each noise realization and apply the FF stacking estimator to each set of five maps before calculating their mean-subtracted power spectra. This Monte Carlo approach captures noise biases beyond our Taylor-approximated expression in Eq. \eqref{eq:ff_corr_Iobs}, e.g., terms $\mathcal{O}(\delta^2[\hat{FF}])$, however these terms are small.


\subsubsection{Multiplicative bias correction}
\label{sec:mode_couple_ff}

In the absence of masking, the FF stacking estimator leads to a multiplicative bias on the sky power spectrum which we derive in \ref{sec:unmasked_ffbias}. For FF weights $w_i$ and sky brightnesses $I_i$,
\begin{equation}
    \frac{\hat{C}_{\ell,j}}{C_{\ell,j}^{true}} \approx 1 + \sum_{i\neq j} \left(\frac{w_i I_j}{I_i}\right)^2. 
\end{equation}
ZL brightnesses across the five \emph{CIBER} \nth{4} flight fields vary by up to a factor of two, meaning the multiplicative bias is of order unity for fields such as elat10. 

Beyond this, an important realization is that the FF stacking estimator couples modes from \emph{all} masked fields with the mask of each target field. This can be seen by writing the masked version of Eq. \eqref{eq:ff_corr_Iobs} (see App. \ref{sec:ffmask}) and modifying the expression for $\delta[\hat{FF}]$. In the limit where $\delta FF \ll 1$, this additional mode coupling can be approximated with linear operators and treated within the standard pseudo-$C_{\ell}$ formalism. Through a modified Monte Carlo procedure (also detailed in the Appendix), we capture the combined mask+FF+filtering mode mixing in a single matrix transformation. This is similar in spirit to \cite{master_leung}, which incorporates the combined mode coupling of time-ordered-data filtering and survey masks into a single matrix transformation $J_{\ell \ell'}$.

\begin{figure*}
    \centering
    \includegraphics[width=0.85\linewidth]{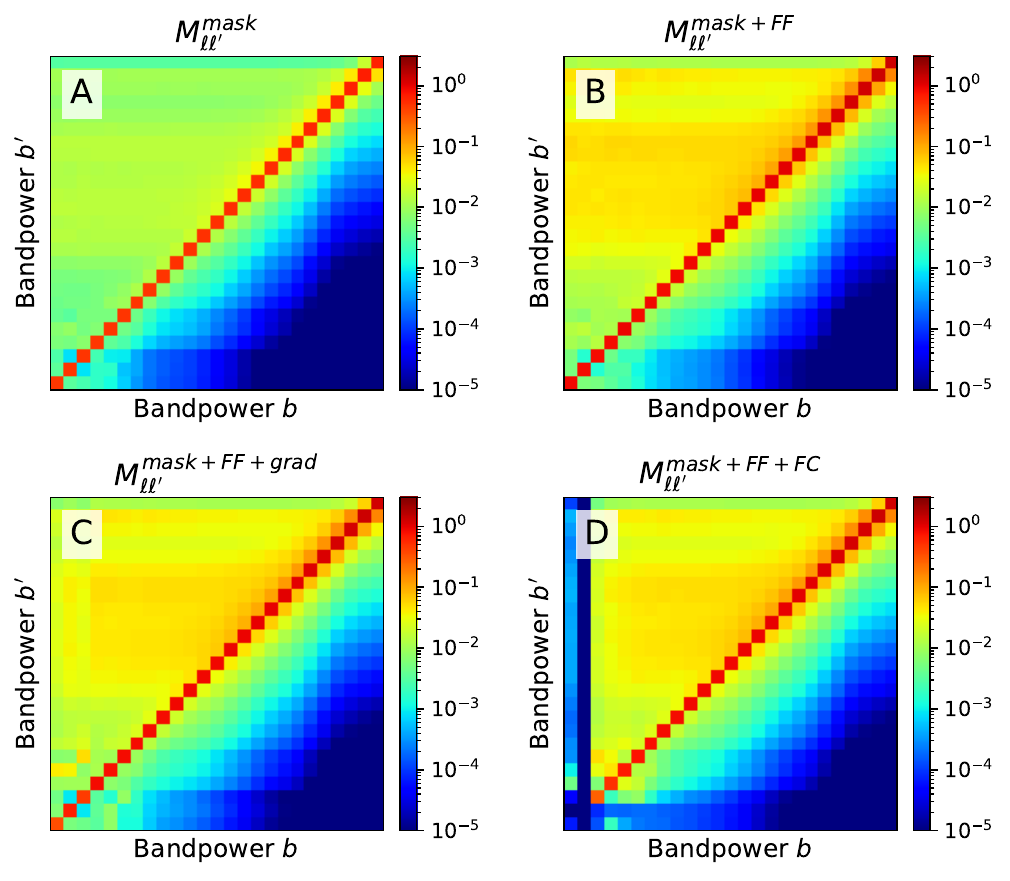}
    \caption{Comparison of mode coupling matrices with and without filtering and other corrections. The standard mode coupling matrix $M_{\ell\ell'}^{mask}$ (case A) derived from the mask is shown in the top left, along with the hybrid mask+FF matrix ($M_{\ell\ell'}^{mask+FF}$, case B). Two versions of the mixing matrix with image filtering are presented in the bottom panel, one using a simple gradient filter (case C) and one with a more aggressive Fourier component filter (case D). In both filtering cases we employ per-quadrant offset fitting. These different mixing matrices highlight the off-diagonal mode couplings induced by our map processing.}
    \label{fig:mkk_vs_mkkff}
\end{figure*}

To compare the effects of mode coupling from different operations on the data, in Fig. \ref{fig:mkk_vs_mkkff} we present several variations on the mixing matrix for a single field. In all configurations, we observe that input power is preferentially redistributed from large to small scales. Compared to the case of mask convolution alone (case A), the mode coupling from cases with FF errors and filtering can be up to a factor of five stronger for certain bandpower combinations. Comparing cases B and C, we see that image filtering modifies the mode coupling structure, primarily for low-$\ell$ bandpowers. Case D highlights the more aggressive \nth{2}-order Fourier component filter, for which low-$\ell$ power is strongly suppressed before it mixes with smaller scales in the pseudo-power spectrum.

Finally, we note that configuration space estimators such as the two-point correlation function are subject to similar additive and multiplicative biases when using the same FF stacking estimator, however they do in principle bypass mode-mixing effects induced by the astronomical masks. We do not pursue a configuration space analysis in this work, however we note its use in existing and near-future diffuse light fluctuation measurements \citep{cappelluti_13, librae}.    



\section{Masking deeper in the NIR with multi-wavelength photometry}

\label{sec:mask}

Having laid out our general pseudo-$C_{\ell}$ formalism, we now turn to a practical analysis challenge, namely effective source masking. For clean measurements of large-angle clustering, source masking reduces the effective shot noise level associated with Poisson fluctuations. While masking deeper removes Poisson fluctuations, there is a trade off with minimizing the fraction of masked pixels. However, by bypassing the need for field differences, we avoid a large penalty in the masking fraction and resulting mode coupling. This is because the masking fraction in field differences is determined by the instrument mask and union of two independent astronomical source masks, meaning uncertainties due to reduction of effective modes and de-convolution of masked power spectra are more severe. Single-field imaging thus enables a more aggressive masking of sources.

Beyond mask de-convolution, effective source masking in \zem\ was limited by external catalog completeness. In \zem, source masks were constructed using $J$-band catalogs from 2MASS to a depth of $J = 17.5$. However, the 2MASS completeness falls quickly beyond $J=16.0$ and $H=15.0$. While deeper NIR catalogs such as UKIDSS and IBIS exist, their coverage within the \emph{CIBER} fields is highly non-uniform and/or unavailable. Another approach would be to over-mask, using sources identified in external catalogs at other wavelengths, e.g., from optical surveys. However the masking would be very inefficient, since it is not clear \emph{a priori} which optically identified sources correspond to the brightest $J$- and $H$-band sources. 

For our approach, we take advantage of the fact that deeper multi-band optical and infrared photometry from PanSTARRS and WISE contain sufficient information to determine cuts on $J$- and $H$-band source magnitudes. Rather than construct hand-crafted color cuts, we make direct $J$- and $H$-band magnitude predictions through random forest regression. We then use these predictions to estimate the infrared flux to set the size of masks surrounding the identified point source. In the following sub-sections we summarize the properties of the external catalogs used to train and test our random forest model.

\subsection{Source catalogs}
In this work we use direct NIR photometry from the 2MASS and UKIDSS catalogs along with optical and infrared photometry from PanSTARRS and WISE. With the exception of UKIDSS, these catalogs have full coverage across all five of the \emph{CIBER} 4th flight science fields.  We summarize the depths of these catalogs in Table \ref{tab:ancillary_catalogs}.

\begin{table}[]
    \centering
    \begin{tabular}{c|c|c}
        Survey & Filters & 5$\sigma$ point source depth \\
        \hline
        PanSTARRS & grizy & 23.4, 23.0, 22.7, 21.8, 20.7\\
        2MASS & JHK & 17.0, 16.3, 15.5 \\
        UKIDSS (LAS)\footnote{Non-uniform/incomplete coverage of elat10, elat30, SWIRE, but not available for the \bootes\ fields.} & JHK & 18.7, 17.4, 16.3\\
        UKIDSS (UDS) & JHK & 24.7, 23.7, 23.4\\
        unWISE & W1 & 17.5
    \end{tabular}
    \caption{List of ancillary catalogs used in this work and their properties. We use 2MASS photometry to mask bright sources in our fields ($J<16$), while for fainter sources we use a combination of PanSTARRS and unWISE photometry to predict NIR magnitudes using a model trained on the UKIDSS UDS catalog (see \S \ref{sec:rftrain_uds}). All listed depths are in the Vega magnitude system.}
    \label{tab:ancillary_catalogs}
\end{table}

\subsubsection{2MASS}
The Two Micron All Sky Survey \citep[2MASS;][]{2MASS} imaged the sky in $J$ (1.2 $\mu$m), $H$ (1.6 $\mu$m) and $K$ (2.1 $\mu$m) bands using 1.3-meter telescopes at Mt. Hopkins and CTIO, Chile. The extended 2MASS catalog is 75\% complete in integrated counts down to $J = 17.5$ (Vega), or 17.5 (18.4 AB) and 17.0 (18.4 AB) for \emph{CIBER}'s 1.1 $\mu$m and 1.8 $\mu$m bands, respectively. In this work 2MASS is used to identify sources with $J < 16$. For very bright sources, the 2MASS point source catalog uses measurements from shorter integrations, either 1.3 second exposures ($\texttt{rdflg}=1$) or 51 ms from the array reset for the brightest sources ($\texttt{rdflg}=3$). Across the five \emph{CIBER} fields (elat10, elat30, \bootes\ B, \bootes\ A and SWIRE), there are $\lbrace 24, 22, 24, 22, 44\rbrace$ stars with $\texttt{rdflg}=1$ while only two stars in elat30 and one in SWIRE have $\texttt{rdflg}=3$.


\subsubsection{Pan-STARRS}

The Panoramic Survey Telescope and Rapid Response System \citep[PanSTARRS;][]{panstarrs} is a system designed for wide-field astronomical imaging . The 1.8-meter telescope, situated on Haleakala in Maui, has a 1.4 Gigapixel camera with 7 deg$^2$ field of view, and has imaged the sky in five broadband filters ($g, r, i, z, y$). The primary 3$\pi$ survey covers 3$\times10^4$ deg$^2$, with full coverage over the \emph{CIBER} fields. We query source positions and magnitudes in these bands from the DR2 \texttt{MeanObject} table, including all sources with $y$-band measurements and quality flags (\texttt{qualityFlag}) in the \texttt{ObjectThin} table equal to 8 or 16. PanSTARRS is a desirable catalog for our purposes given its relatively deep $y$-band photometry, which more strongly correlates with $J$- and $H$-band fluxes.

\subsubsection{unWISE}

The unWISE catalog consists of photometry from unblurred coadds from WISE imaging \citep{unWISE, Schlafly18}. We use the five-year catalog, which at 3.4 $\mu$m has a $5\sigma$ depth of $W1=17.5$ ($W1_{AB}=20.8$). The performance using both $W1$ and $W2$ photometry was comparable to that from $W1$ alone, so we opt for the latter.

\subsubsection{UKIDSS}
The UKIRT Infrared Deep Sky Survey \citep[UKIDSS;][]{ukidss} consists of seven years of imaging in the near-infrared with varying depths, carried out using the UKIRT Wide Field Camera (WFCAM). The deepest coverage available covers 0.77 deg$^2$ in the Ultra Deep Survey (UDS) field down to $J=24.7$, $H=23.7$ and $K_s=23.4$. We use the UKIDSS UDS photometry to train our random forest regression method, which is detailed in the next sub-section. The UKIDSS Large Area Survey (LAS) is considerably shallower than UDS (depth of $K\sim 18$) but has available $J$- and $H$-band photometry for two of the five \emph{CIBER} fields (elat10 and elat30). Lastly, the UKIDSS Deep Extragalactic Survey (DXS) reached a depth of $K\sim 21$ and covers the majority of the SWIRE field for $J$ band. The LAS and DXS photometry are compared against our predicted catalogs where available.

\subsection{Random forest model and training}
\label{sec:rftrain_uds}
We perform random forest regression training and validation on UKIDSS photometry in the UDS field, which probes significantly deeper than our desired masking depths. We cross-match the UDS catalog against unWISE and PanSTARRS using a matching radius of 1\arcsec. When a source lacks a PanSTARRS or unWISE detection, the missing magnitudes are replaced with $m=30$, i.e., they are labeled as non-detections.

We split 70\% and 30\% of the cross-matched catalog to form our training and validation samples, respectively. We restrict our samples to sources with $J<21$ to prevent training set imbalance relative to our target \emph{CIBER} masking depths. We use the publicly available package \texttt{sklearn} to train separate random forest models for $J$- and $H$-band predictions. We set the maximum depth of the decision trees to eight, beyond which the regression performance plateaus.

\begin{figure*}
    \centering
    \includegraphics[width=0.61\linewidth]{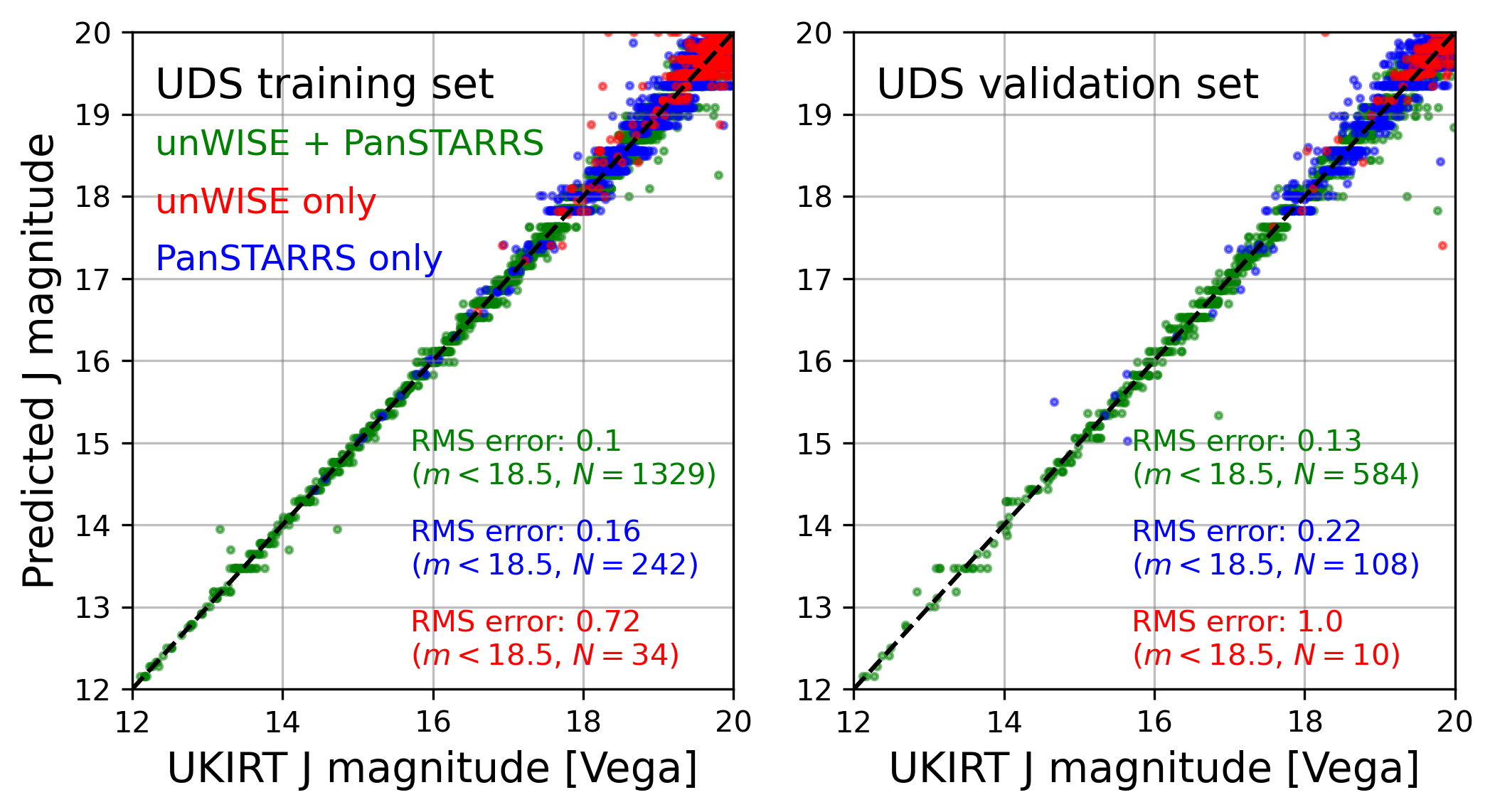}
    \includegraphics[width=0.33\linewidth]{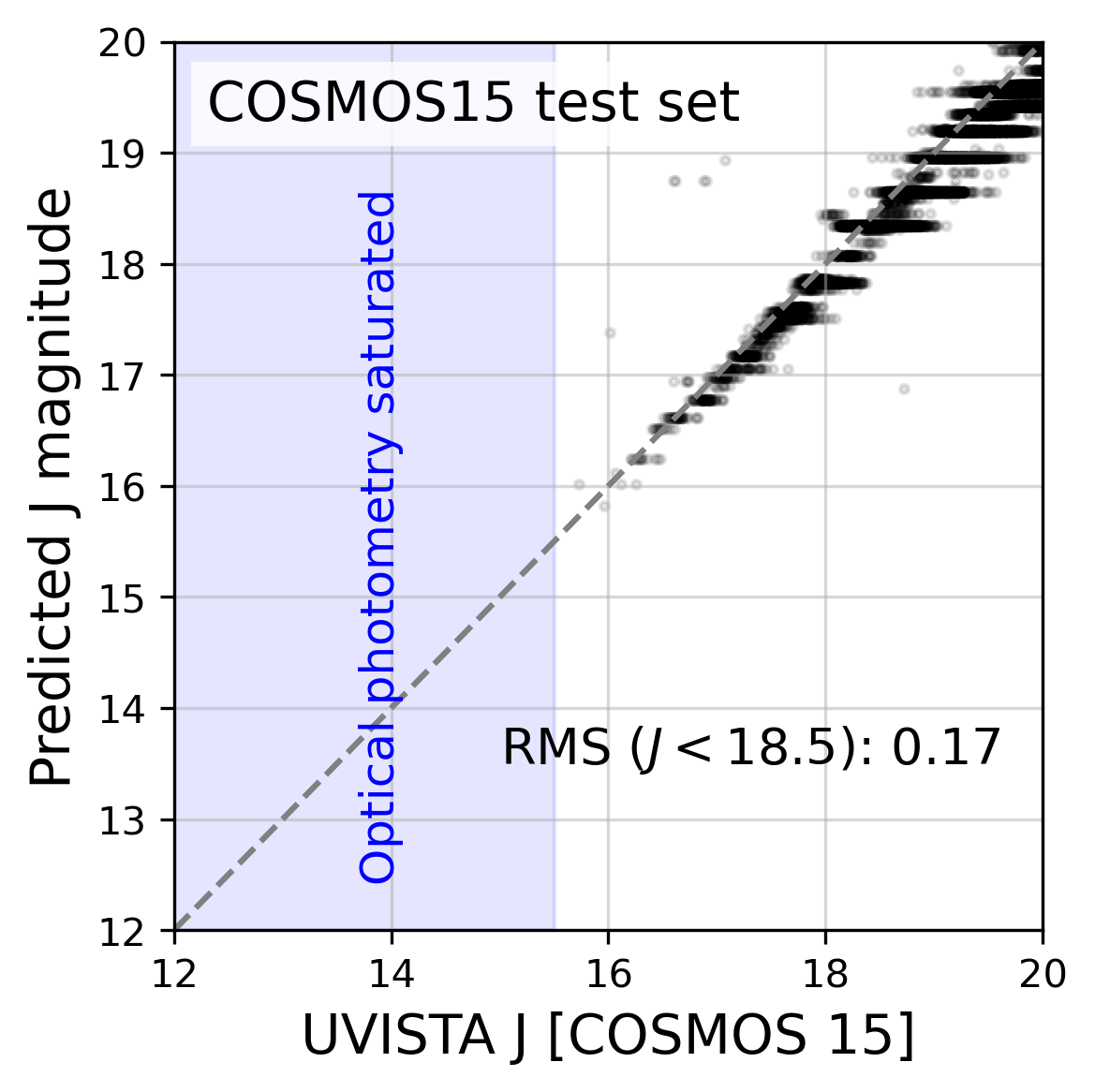}    
    \includegraphics[width=0.61\linewidth]{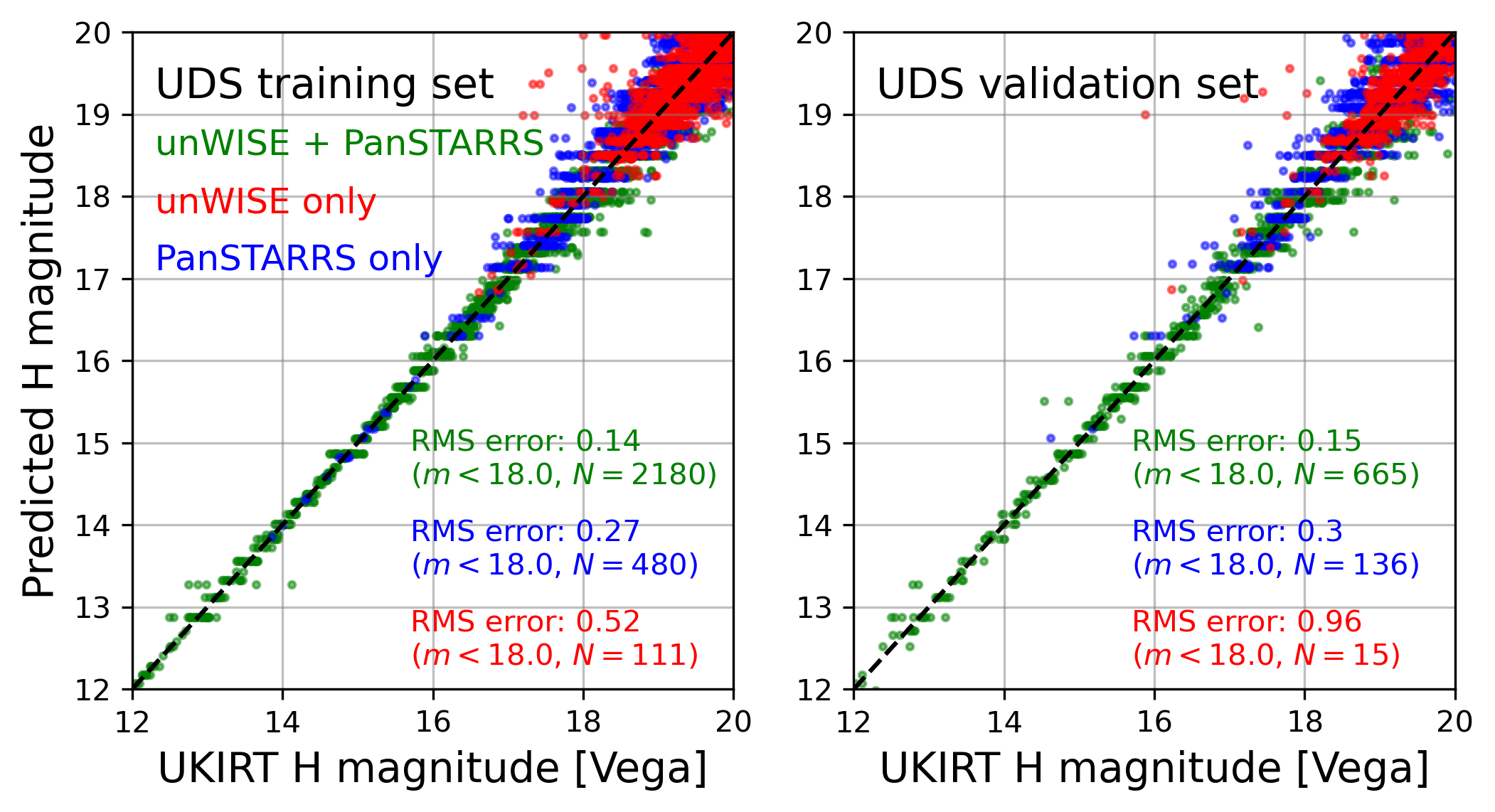}
    \includegraphics[width=0.33\linewidth]{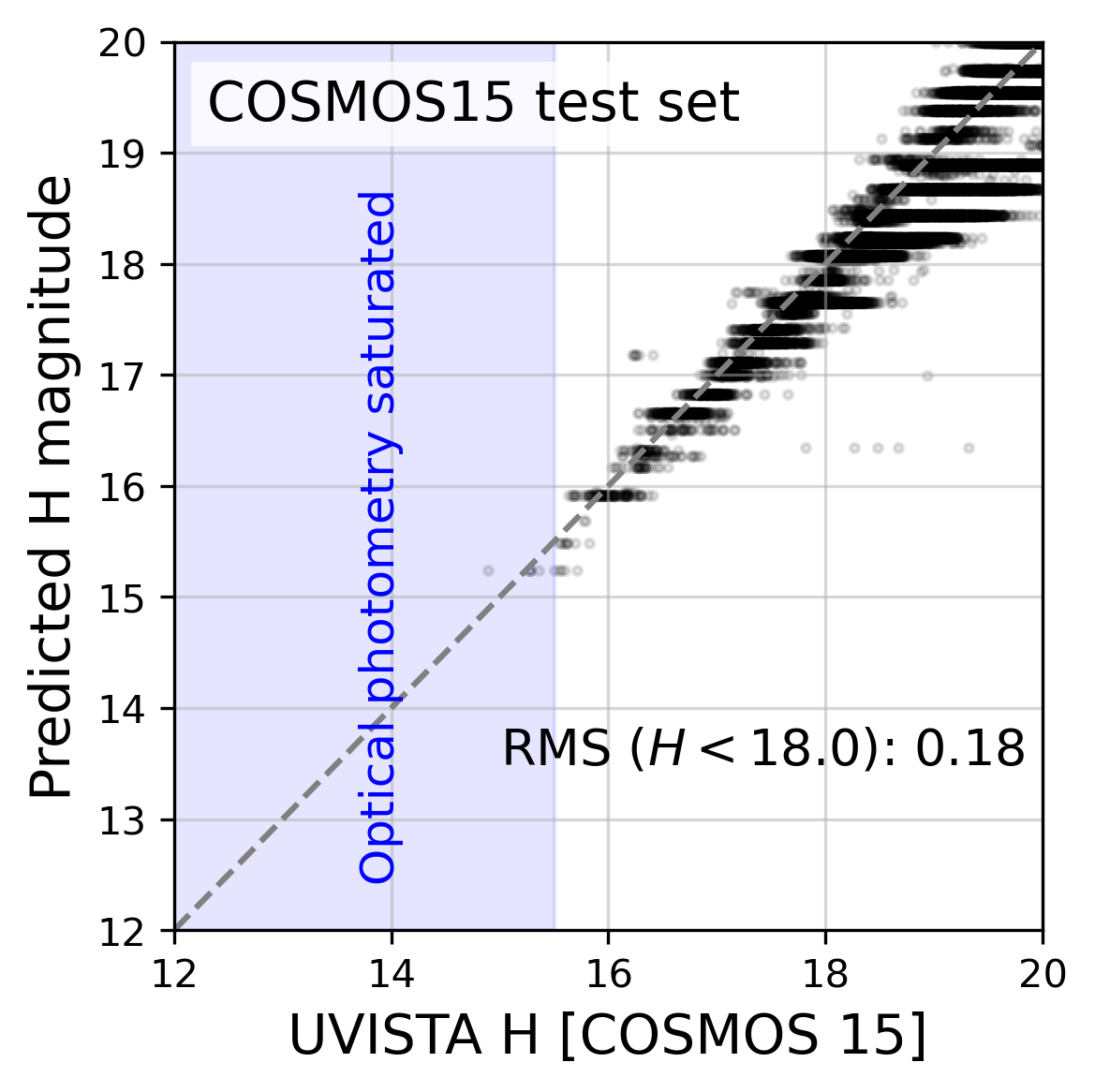}    
    \caption{Comparison of measured magnitudes and random forest-predicted magnitudes using ancillary photometry, for $J$-band (top row) and $H$-band (bottom). The left and middle columns show our results for UKIDSS training and validation sets, respectively. UKIDSS sources with both unWISE and PanSTARRS counterparts are plotted in green, while those with only unWISE or only PanSTARRS are plotted in red and blue, respectively. The right-hand column shows the results of applying our models trained on UKIDSS to COSMOS 2015 photometry, as a test of distribution shift. The COSMOS field used Suprime-Cam and IRAC for the optical and IR data.
    }
    \label{fig:dt_train_validate}
\end{figure*}

In Figure \ref{fig:dt_train_validate} we compare the predicted $J$- and $H$-band magnitudes with those from the UKIDSS training and validation sets. The results are unbiased on average, with increasing dispersion toward fainter magnitudes. There is a small number of catastrophic outliers, which typically correspond to sources with incomplete multi-band coverage, e.g., unWISE only (red points) or PanSTARRS only (blue points). The error RMS for the sub-samples, indicated in Fig. \ref{fig:dt_train_validate}, are smallest when both optical and infrared photometry are available and largest when only infrared photometry is available. 

We summarize the completeness and purity of the derived UDS masking catalogs in Table \ref{tab:dt_completeness_purity} for a range of masking thresholds. The total number of predicted sources below each magnitude threshold agrees well with those from our validation samples and the completeness and purity of our predicted catalog is $>90\%$ in all cases, with some degradation toward fainter fluxes. As expected, masking predictions for sources with both optical and infrared counterparts have the highest completeness and purity. In comparison, the completeness of 2MASS catalog in the same field falls off quickly, starting at 89\% and 82\% for $J<17.5$ and $H<17.0$, respectively, and going down to 34\% and 27\% for $J<19.0$ and $H<18.5$. While incompleteness in our masking catalog leads to additional point source fluctuation power, the main penalty of catalog impurity is a slightly higher masking fraction.


\begin{table*}
    \centering
    \begin{tabular}{c|c|c|c|c|c|c|c}
        Masking depth & N$_{src}$ & N$_{src}$, $\mathcal{C}$, $\mathcal{P}$ & $\mathcal{C}$, $\mathcal{P}$ & $\mathcal{C}$, $\mathcal{P}$ & $\mathcal{C}$, $\mathcal{P}$ & $\mathcal{C}$, $\mathcal{P}$ & $\delta C_{\ell}^{SN}/C_{\ell}^{SN}$ \\
        Vega & (UKIDSS) & all, predicted & PS+unWISE & PS only & unWISE only & 2MASS only & C15 test set \\
         \hline 
         \hline
        $J<17.5$ & 370 & (371, 0.98, 0.98) & (0.98, 0.98) & (0.90, 1.0) & (0.0, 0.0)\footnote{There is only one source which satisfies this condition in our validation set.} & (0.89, -) & -5.2\%\\
        $J<18.0$ & 498 & (481, 0.93, 0.97) & (0.95, 0.98) & (0.88, 0.87) & (0.16, 0.33)& (0.65, -) & -10.2\%\\
        $J<18.5$ & 684 & (640, 0.92, 0.98) & (0.95, 0.99) & (0.82, 0.97) & (0.57, 0.88)& (0.47, -) & -12.2\%\\
        $J<19.0$ & 977 & (947, 0.93, 0.95) & (0.96, 0.96) & (0.88, 0.92) & (0.45, 0.90)& (0.34, -) & -21.7\%\\
        \hline
        $H<17.0$ & 422 & (416, 0.96, 0.97) & (0.97, 0.98) & (0.79, 0.85) & (0.33, 0.25) & (0.82, -) & -6.9\%\\
        $H<17.5$ & 574 & (544, 0.91, 0.97) & (0.93, 0.98)& (0.81, 0.83) & (0.33, 0.60) & (0.58, -) & -2.2\%\\
        $H<18.0$ & 828 & (829, 0.94, 0.94) & (0.98, 0.94) & (0.73, 0.92) & (0.83, 0.68) & (0.40, -) & -8.8\%\\
        $H<18.5$ & 1206 & (1167, 0.90, 0.93)& (0.94, 0.95) & (0.84, 0.84) &  (0.54, 0.73) & (0.27, -) & -16.4\% \\
    \end{tabular}
    \caption{Random forest regression performance on our UDS validation set, for various masking selections. The table shows the total number of sources brighter than each masking threshold from UKIDSS, along with the completeness ($\mathcal{C}$) and purity ($\mathcal{P}$) of the predicted catalogs. We also include the mean fractional power spectrum bias on Poisson fluctuations ($\ell > 10000$), which we estimated by applying our pre-trained model on UDS to the COSMOS 2015 catalog and computing the true vs. estimated fluctuation power of sources fainter than each magnitude cut. The negative signs indicate that the masking method removes more Poisson fluctuations than an ideal mask at the specified $J$- or $H$-band magnitude.}
    \label{tab:dt_completeness_purity}
\end{table*}

\subsection{Testing mask predictions with COSMOS}

To assess any systematic uncertainties due to distribution shift between our UDS training set and the science fields, we apply the model to multi-band photometry from the COSMOS 2015 catalog \citep{laigle15}. One subtlety is that the optical and infrared photometry in COSMOS come from Suprime-Cam and IRAC rather than PanSTARRS/WISE, adding a layer of distribution shift beyond our application in the science fields. Nonetheless we use our pretrained model to predict $J$- and $H$-band magnitudes and compare these against measured magnitudes from the COSMOS catalog. 

We include the distribution of predicted and measured magnitudes in the right column of Fig. \ref{fig:dt_train_validate}. Due to saturation in some optical bands for the COSMOS catalog, our results are limited to $J>16$, which is our main focus in any case. Our predictions match the COSMOS15 measured magnitudes closely for a range of fluxes, however there is larger scatter and a mild negative bias on the predicted magnitudes. For sources down to $J<18.5$ and $H<17.0$, the error RMS for each band is 20\%/30\% higher than that of the UDS validation sets, which corresponds to lower purity in the test set results. This may be due to differences in source photometry across catalogs or calibration discrepancies between PanSTARRS/Suprime-Cam and WISE/IRAC. COSMOS15 is more complete in the optical/IR than the PanSTARRS/WISE catalog, leading to well-determined photometry for some sources that would be otherwise labeled as non-detections in the training set. 

Given our predicted masking selections, we then calculate the sub-threshold Poisson noise of unmasked sources and compare against the ``true" Poisson noise at fixed masking depth. The fractional shot noise errors are included in Table \ref{tab:dt_completeness_purity}. Our results suggest a slight over-removal of point source power in the predicted catalogs, though the fractional difference in power is small ($<13\%$ for $J < 18.5$ and $<9\%$ for $H < 18.0$). For the deepest masking depths ($J<19.0$ and $H<18.5$) the departures are slightly larger ($22\%$ and $16\%$ for $J$- and $H$-band respectively). A full interpretation of these discrepancies needs to take into account systematic differences between the training and test sets, however in general these results suggest that our source masking procedure is robust, extending over two Vega magnitudes deeper than through 2MASS alone.



\subsubsection{Recovered source counts}

\begin{figure*}
    \centering
    \includegraphics[width=0.9\linewidth]{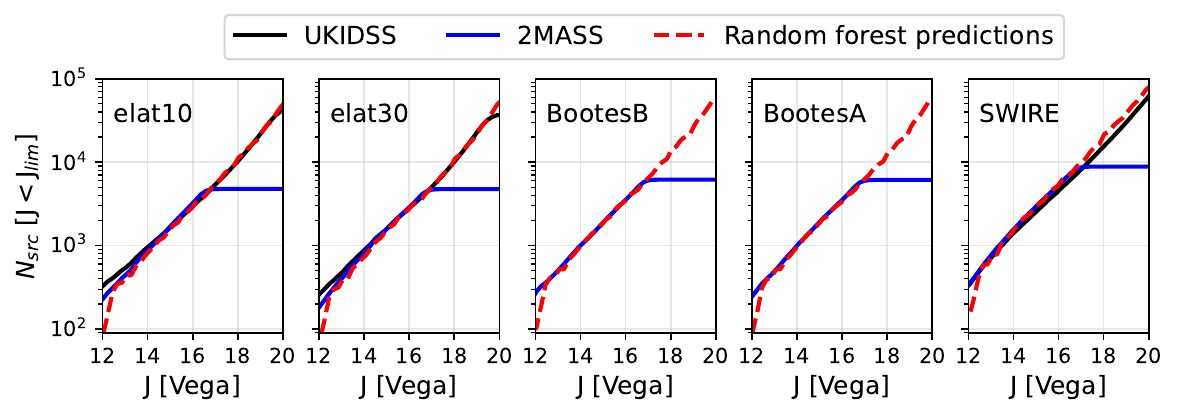}   
    \includegraphics[width=0.9\linewidth]{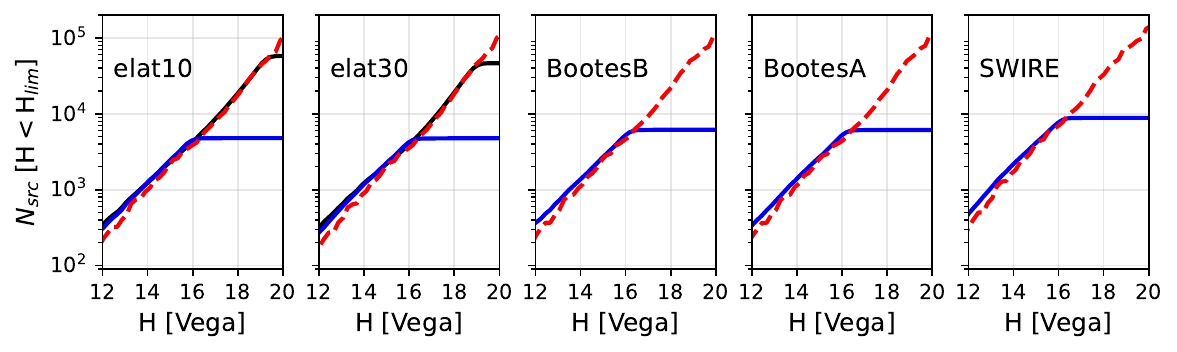}   
    \caption{Cumulative number counts from 2MASS (blue), UKIDSS (black) and random forest-predicted magnitudes using PanSTARRS+unWISE photometry (red). Our predicted catalogs extend several magnitudes beyond 2MASS and have consistent number density to UKIDSS Large Area Survey (LAS) counts available in two of the five \emph{CIBER} fields (elat10, elat30). The SWIRE field is covered by the UKIDSS Deep Extragalactic Survey (DXS) for $J$ band, for which our random forest predicts slightly higher counts.}
\label{fig:mag_dist_compare_catalogs}
\end{figure*}

Figure \ref{fig:mag_dist_compare_catalogs} shows the $J$- and $H$-band cumulative number counts in the five \emph{CIBER} science fields, recovered from different catalogs. We limit our comparisons to $J>12$, the brightest magnitude available in our UDS catalog. The 2MASS catalogs are in broad agreement with both our predictions and UKIDSS down to $J=16.5$ and $H=16$. Beyond these depths the 2MASS catalog becomes incomplete, which we quantify in Table \ref{tab:dt_completeness_purity}. In both bands, our predicted counts for the elat10 and elat30 fields are consistent with UKIDSS LAS to within $5\%$ for $J<18.5$ and $H<18$. In the SWIRE field, we see larger differences, with our predicted integrated counts higher than UKIDSS by 20-35\% going from $J=16$ to $J=18.5$. 

For the brightest sources ($12\leq m \leq 15$), our predicted catalogs exhibit larger discrepancies with respect to 2MASS and UKIDSS. We attribute this to the small training set of bright sources within the UDS field. Rather than develop a larger bright-end training set for our random forest model, we simply merge the bright end of the 2MASS catalog ($J<16$ and $H<15$) with our random forest-derived catalogs to obtain our final masking catalogs.

\subsubsection{Masking catalog consistency with simulations}

\begin{figure*}
    \centering
    \includegraphics[width=\linewidth]{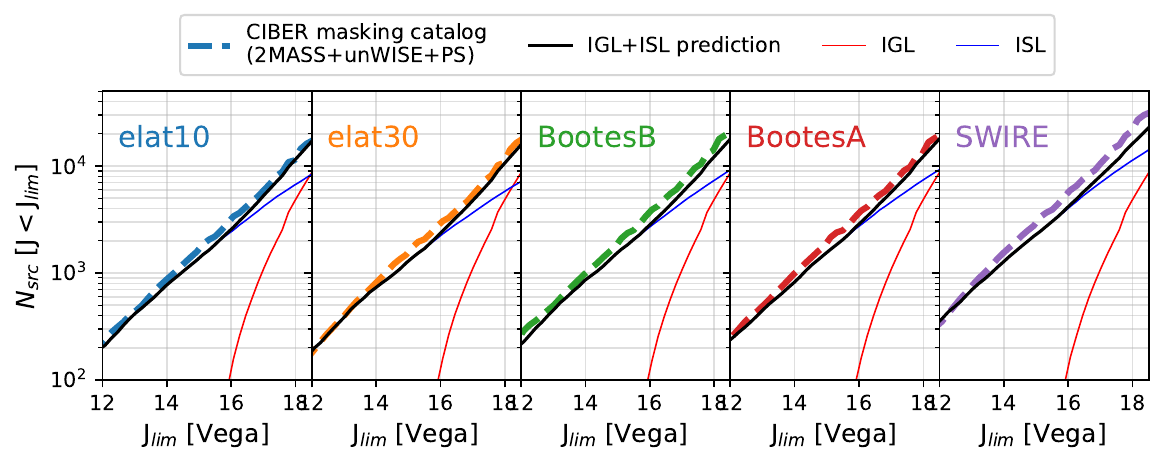}
    \includegraphics[width=\linewidth]{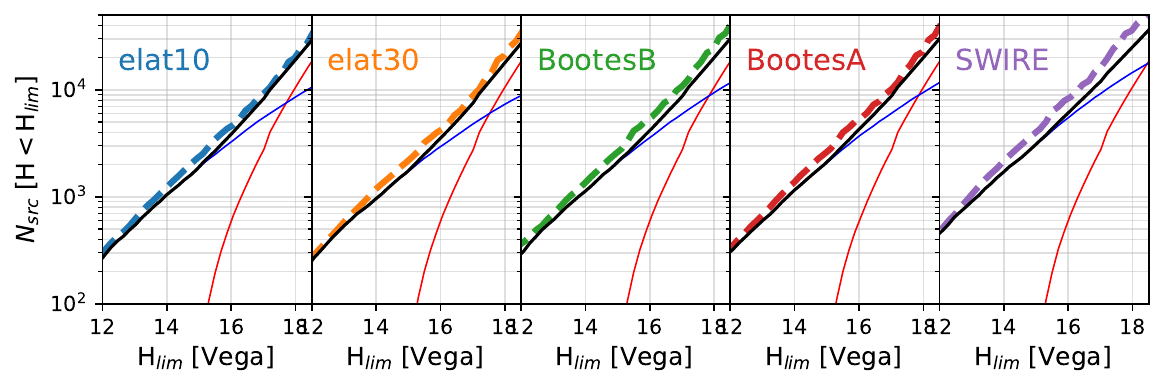}
    \caption{Cumulative magnitude distributions of the final \emph{CIBER} masking catalogs (dashed lines) for $J$-band (top) and $H$-band (bottom), compared with those derived from IGL+ISL predictions for the same fields (black). Also shown are the cumulative magnitude distributions for stars (ISL, blue) and galaxies (IGL, red) that comprise our predictions.}
    \label{fig:mag_cdf_compare}
\end{figure*}

In Figure \ref{fig:mag_cdf_compare} we compare the cumulative magnitude distributions of our final masking catalogs with those from predicted from simulations. The simulated catalogs combine the TRILEGAL stellar model for each field with realizations of the \cite{helgason} galaxy model. We find that for $16 < m < 18.5$, our final counts are slightly higher than simulations. For the four non-SWIRE fields the counts are consistently higher by 10-20\%, while for SWIRE our counts are 30-40\% higher. The discrepancy with SWIRE is of similar magnitude to that seen between our predicted catalogs and UKIDSS DXS for $J$ band, suggesting potential errors in our predicted catalog. The counts of our IGL mocks are constrained by the \cite{helgason} best-fit luminosity functions, which appear to underestimate measured counts in this magnitude range (c.f. Fig. 12 of \cite{helgason}).

\subsection{Source mask radius prescription}
To construct the \emph{CIBER} astronomical masks we model the masking radius for each source as a function of magnitude. For bright sources ($J<14$ or $H < 14$), we parameterize the masking function as a function of magnitude $m$
\begin{equation}
    r(m) [\arcsec] = A\exp\left[-\frac{(m - b)^2}{c^2} \right],
    \label{eq:mask_radius}
\end{equation}
where $b=3.6$, $c=8.5$ and $A=160$ for all fields. For fainter sources the masking radius is determined iteratively. For each magnitude bin with $\Delta m=0.5$ in the range $14 < J < 19$, we generate a model image of sources in that bin using the measured \emph{CIBER} PSF. The masking radius for those sources is increased until the masked image has power $C_{\ell}^{ePSF}<10^{-9}$ nW$^2$ m$^{-4}$ sr$^{-1}$ for all bandpowers. We perform the bin-wise approach for each field separately in order to capture variations such as PSF size and stellar density. We set a minimum masking radius of 1.5 pixels (10.5\arcsec) for all sources. The fact that the \emph{CIBER} PSF is undersampled means most of the flux for faint sources can be masked across a few pixels. This masking prescription is more aggressive for bright sources than \zem. 

\begin{figure}
    \centering 
    \includegraphics[width=\linewidth]{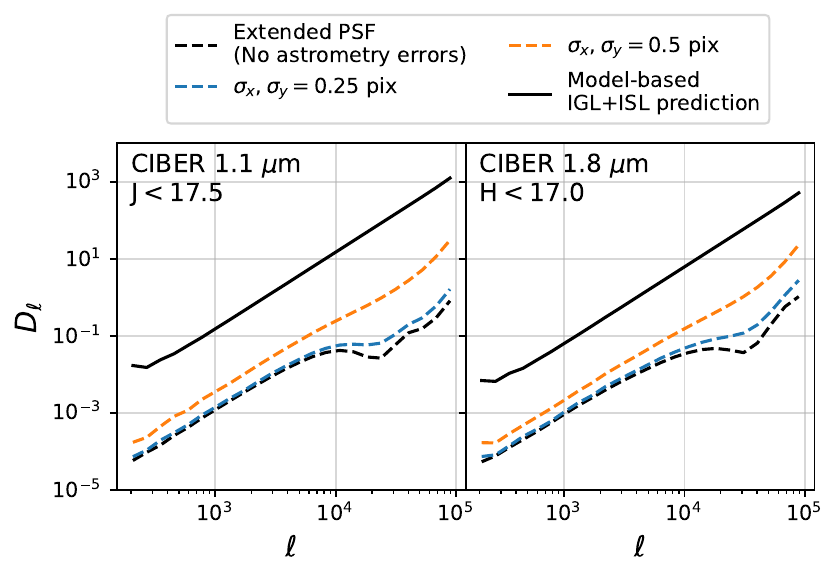}
    \caption{Comparison of power from mask halos in each field (dashed curves) and sub-threshold ISL and IGL fluctuations (solid black line). These are obtained from 100 realizations per field and their respective source masks. We show the case of no astrometry errors (black dashed lines), while results for non-zero astrometry errors are shown in blue and orange.}
    \label{fig:epsf_mocks}
\end{figure}

We then combine the resulting source masks with the \emph{CIBER} 4th flight instrument masks and use these to estimate the total contribution from extended PSF of masked sources. Specifically, for each realization we generate a map of sources down to our fiducial masking depth ($J=17.5$ and $H=17.0$ for 1.1 $\mu$m and 1.8 $\mu$m, respectively), apply its corresponding mask, and calculate the resulting power spectrum. We then correct for the effects of mode coupling and the beam transfer function. We show the results of this exercise performed on 100 sets of mocks in Fig. \ref{fig:epsf_mocks}. In the limit with no astrometric errors, the residual power from masked source halos is more than two orders of magnitude below the IGL+ISL signal and can be considered negligible. To simulate astrometry errors in the \emph{CIBER} pointing solution, we perturb the positions of the injected sources in each dimension by $\sigma_x = 0.25$ and 0.5 \emph{CIBER} pixels (1.75\arcsec\ and 3.5\arcsec\ respectively). In the presence of these astrometry errors the residual power from masked sources increases most significantly at $\ell > 10000$, however the residual power is still much lower than the IGL+ISL signal.

\section{Mock power spectrum recovery tests}
\label{sec:mock_ps_test}

Using the mock \emph{CIBER} observations described in \S \ref{Sec:mocks} and the PS formalism from \S \ref{Sec:ff_bias_formalism} we test our ability to recover sky fluctuations.  Running the pipeline as implemented on several independent mock observations helps to identify any biases in the PS estimation and to quantify measurement uncertainties. For each test configuration we run our pipeline on one thousand sets of \emph{CIBER} mocks, estimating the power spectrum in twenty-five logarithmically spaced bandpowers. In these mock tests we assume perfect knowledge for source masking, i.e., we do not directly emulate masking errors.

\subsection{Field-averaged power spectrum}
To optimally combine power spectrum estimates from the five \emph{CIBER} fields we apply per-bandpower inverse variance weights derived from the dispersion of recovered mock power spectra. We show these weights as a function of multipole in Appendix \ref{sec:field_weights_app} for mocks with FF errors. While on large angular scales the weights are relatively consistent with uncertainties driven by sample variance, differences in read noise and photon noise across the five fields drive a larger dispersion in the bandpower weights on intermediate and small scales.

\subsection{Effect of flat field errors}

\begin{figure*}
    \centering 
    \includegraphics[width=0.47\linewidth]{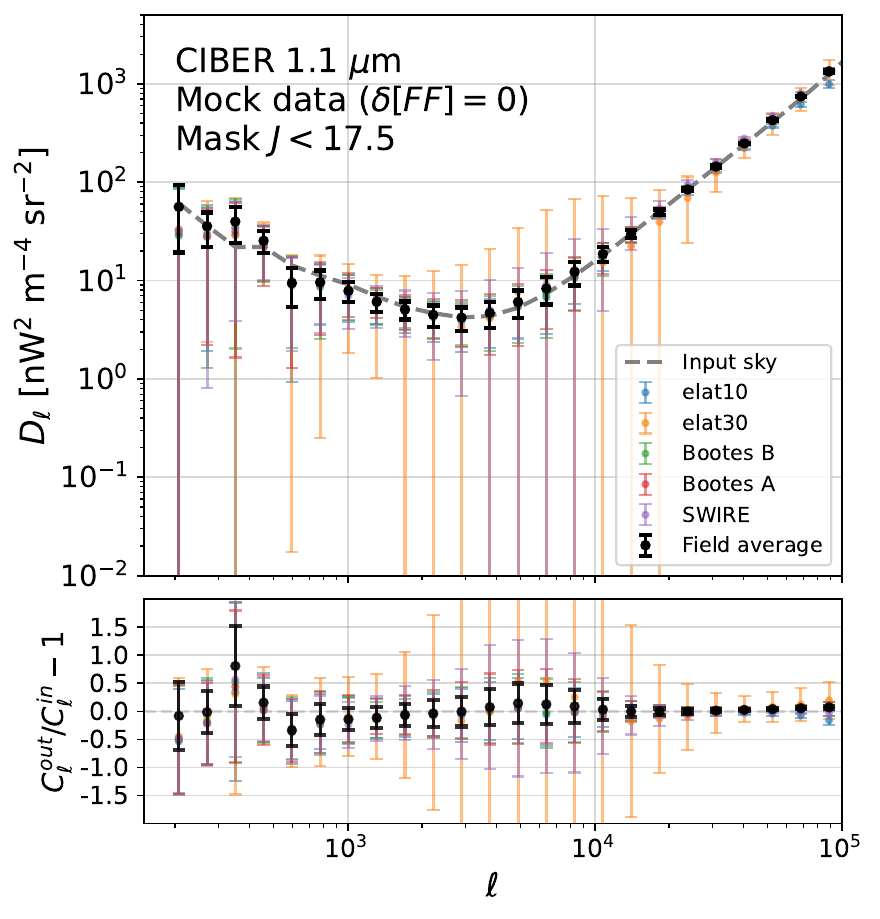}
    \includegraphics[width=0.47\linewidth]{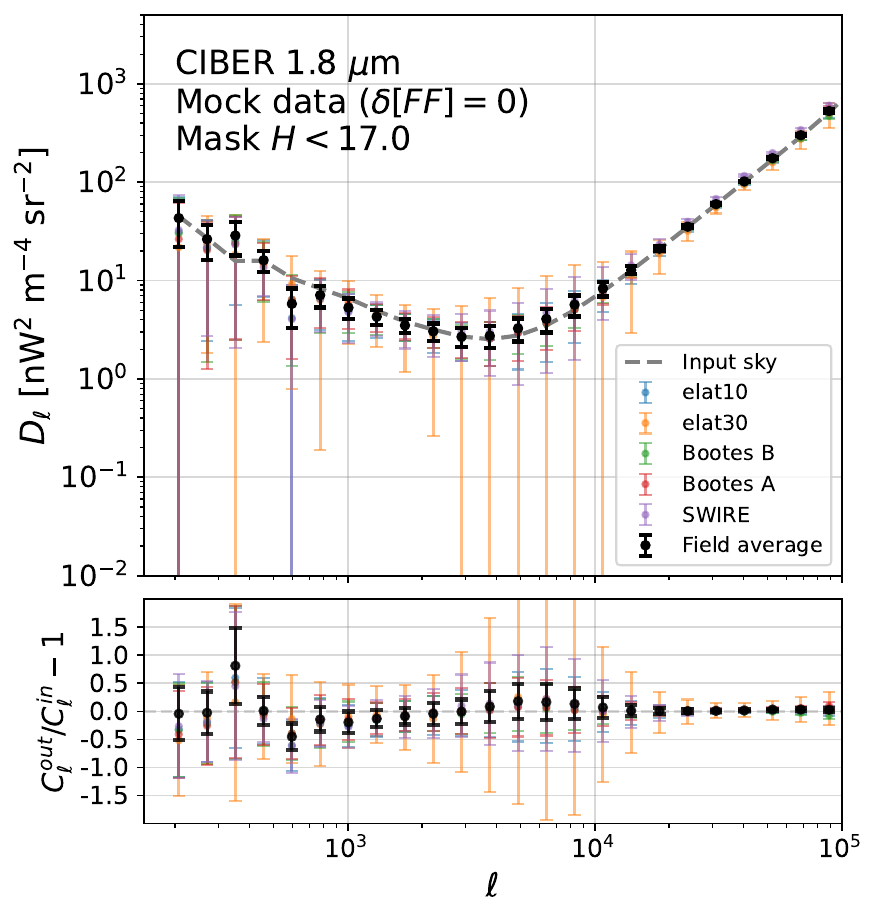}
    \caption{Mock power spectrum recovery with no FF errors ($\delta[\hat{FF}]=0$), for individual fields (colored points) and field averages (black points), plotted for 1.1 $\mu$m (left) and 1.8 $\mu$m (right). The errorbars on the black points are computed from the mean and dispersion of recovered power spectra from one thousand sets of mocks, where each set denotes a realization of five \emph{CIBER} fields. The bottom row shows the fractional power spectrum error relative to the input power spectra.}
    \label{fig:mock_recover_perfectFF}
\end{figure*}

\begin{figure*}
    \centering
    \includegraphics[width=0.47\linewidth]{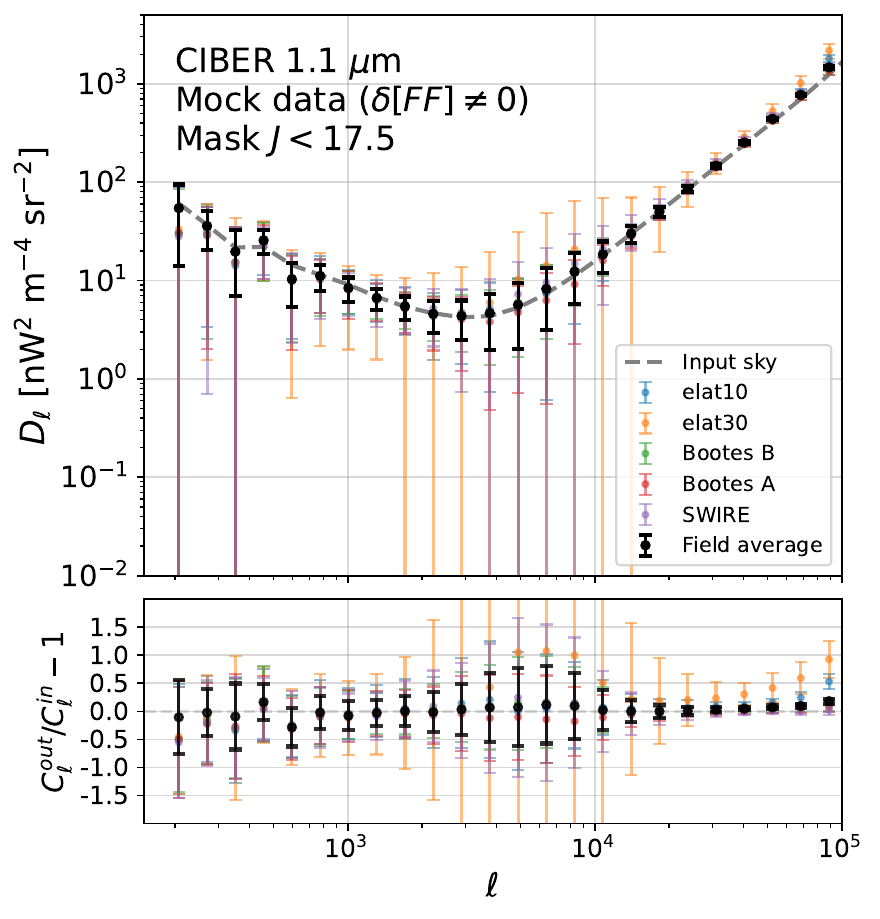}
    \includegraphics[width=0.47\linewidth]{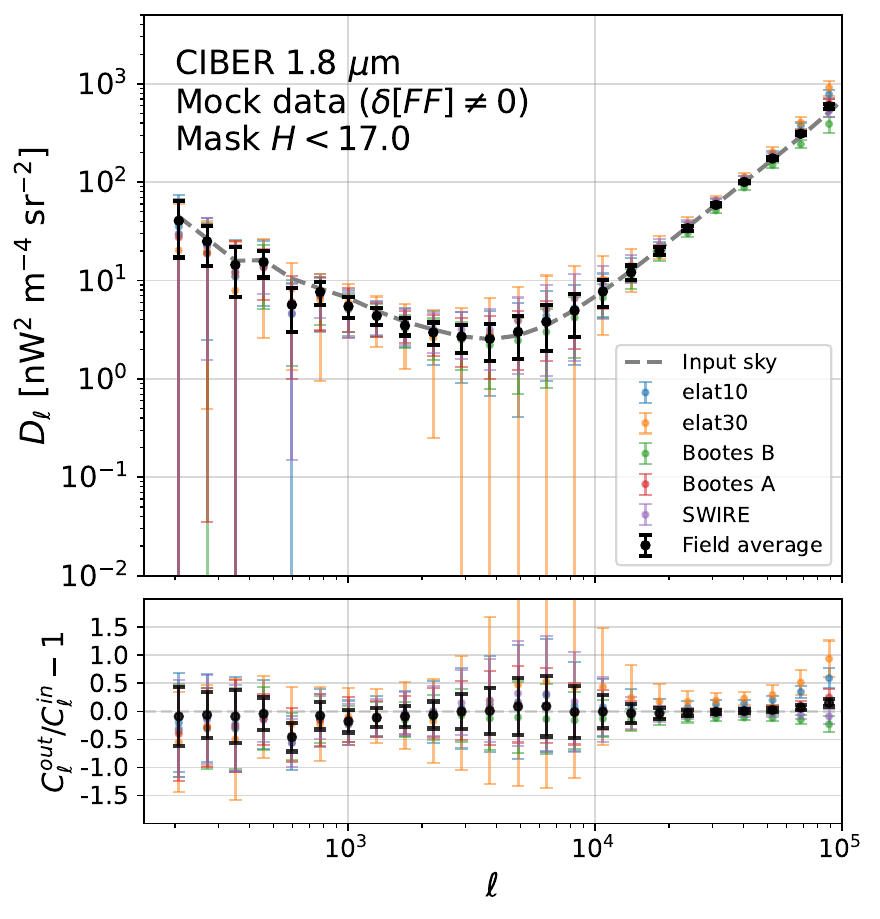}
    \caption{Mock power spectrum recovery with estimated FFs ($\delta[\hat{FF}]\neq 0$) using the stacking estimator from \S \ref{Sec:ff_bias_formalism}. In these tests we use laboratory FF templates (see Fig. \ref{fig:lab_ff}) when generating mock \emph{CIBER} observations.}
    \label{fig:mock_recover_estFF}
\end{figure*}

We validate our pipeline with two test cases at the fiducial masking depths from \zem, namely $J<17.5$ for 1.1 $\mu$m and $H<17.0$ for 1.8 $\mu$m. The first case assumes perfect knowledge of the FF (i.e., no FF correction is needed), while the second incorporates the FF estimation and bias corrections. In Figures \ref{fig:mock_recover_perfectFF} and \ref{fig:mock_recover_estFF} we show the results of these tests. In both cases, the recovered power spectra for elat30 (orange) are much noisier than the other fields, due to the field's short exposure time (17.8 seconds compared to $\sim 50$ for the other science fields). On intermediate scales, the SWIRE field (purple) has large uncertainties due to its higher stellar density and thus masking fraction, despite having the lowest photon noise. 

Averaged over the ensemble of mocks, our per-field and averaged power spectra are unbiased on large and small scales, with some exceptions. On large scales in all cases, the fifth bandpower is negatively biased at the $1-2\sigma$ level in both the $\delta[\hat{FF}]=0$ and $\delta[\hat{FF}]\neq 0$ cases. We attribute this to effects of strong mode coupling between the low-$\ell$ bandpowers introduced by our image filtering (see bottom right panel of Fig. \ref{fig:mkk_vs_mkkff}). In the $\delta[\hat{FF}]=0$ case we also find a $\sim 1\sigma$ positive bias in the third bandpower for both bands, however this is not seen in the  The bias is not $\delta[\hat{FF}]\neq 0$ case. On scales $\ell > 50000$ in the $\delta[\hat{FF}]\neq 0$ case, we find a slight positive bias. We do not find this in the $\delta[\hat{FF}]=0$ case, which suggests some instability in the FF noise bias correction. This is corroborated by the fact that the power spectrum bias is largest for fields elat10 and elat30, which are most sensitive to errors in the noise bias correction due to their high mean sky intensities that couple to instrumental errors in the FF correction.

\begin{figure}
    \centering 
\includegraphics[width=\linewidth]{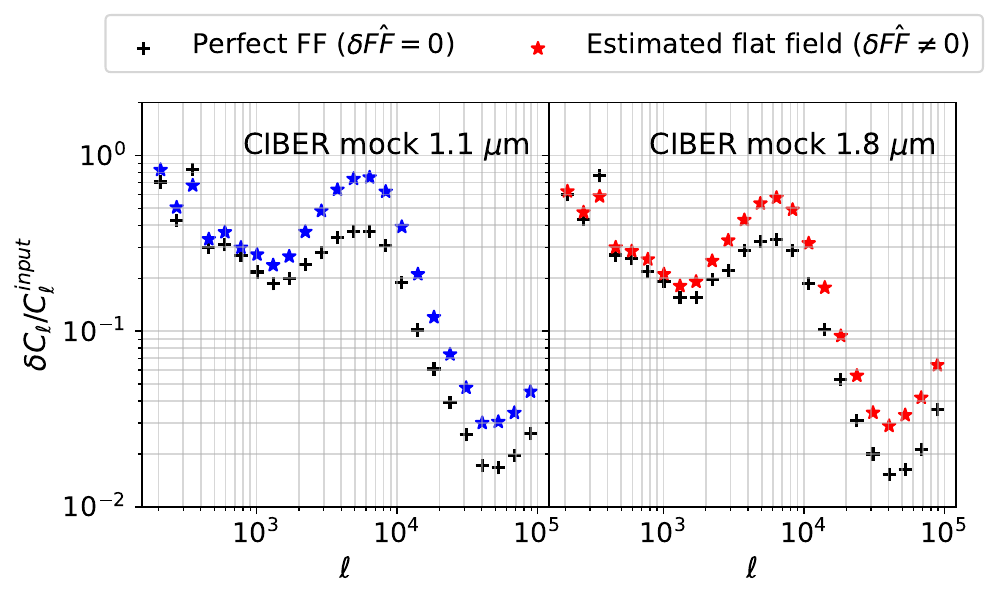}
\caption{Fractional power spectrum uncertainties at 1.1 $\mu$m (left) and 1.8 $\mu$m (right) derived from the dispersion of recovered mock power spectra. We indicate results with and without FF errors using stars and crosses respectively.}
\label{fig:frac_ps_error_mock_compare}
\end{figure}

To understand the sensitivity of our power spectrum measurements as a function of angular scale, in Fig. \ref{fig:frac_ps_error_mock_compare} we plot the fractional power spectrum uncertainties for both bands and test cases. On scales $\ell < 1000$, the uncertainties are driven by sample variance and the filtering transfer function. Our uncertainties peak again on intermediate scales near $\ell \sim 6000$, the result of concentrated power from read noise that spreads to other modes through the mask(s). On small scales, we are dominated by noise, due to the exponential dependence of the beam correction. These results suggest the large-scale sensitivity peaks between $1000<\ell< 2000$, corresponding to angular scales $5\arcmin < \theta < 10\arcmin$.

\begin{figure}
\centering    
\includegraphics[width=0.95\linewidth]{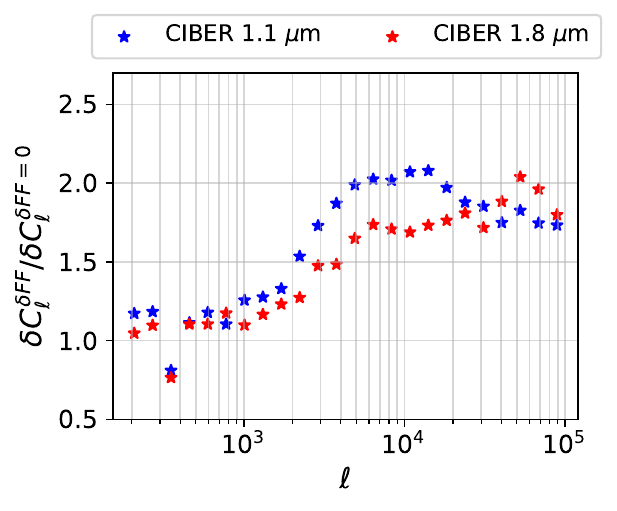}
    \caption{Ratio of power spectrum uncertainties in recovery with and without FF errors. While there is a clear degradation in sensitivity on scales $\ell > 3000$, the increased uncertainty on large scales is relatively modest.}
    \label{fig:relative_frac_ps_error}
\end{figure}

These mock recovery tests enable us to isolate the impact of FF errors on our final measurements. We highlight this in Fig. \ref{fig:relative_frac_ps_error}, plotting the ratio of power spectrum uncertainties between test cases. While the uncertainties are consistently larger in the presence of FF errors as one would expect, we do find an exception in the third lowest bandpower ($\ell \sim 350$) for both bands. This is the same bandpower for which we find a 1$\sigma$ bias and likewise may be explained by the mode couplings induced by FF errors. The degradation in sensitivity is generally larger for 1.1 $\mu$m than 1.8 $\mu$m due to higher photon and read noise levels. Fortuitously, the degradation in sensitivity due to FF errors is modest on large scales, remaining at the $<20\%$ level for $\ell < 1000$ and $20-30\%$ for $1000 < \ell < 2000$.

\subsection{Bandpower correlations}
\begin{figure*}
    \centering \includegraphics[width=0.48\linewidth]{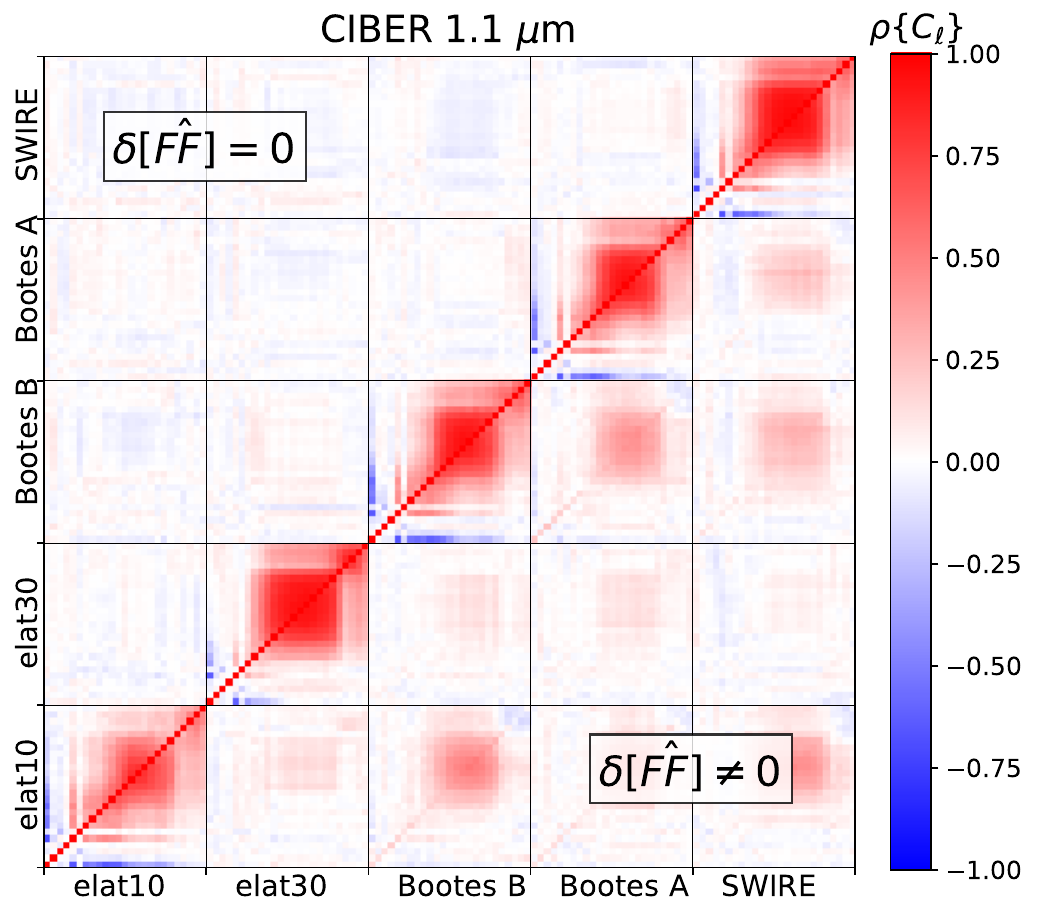}
    \includegraphics[width=0.48\linewidth]{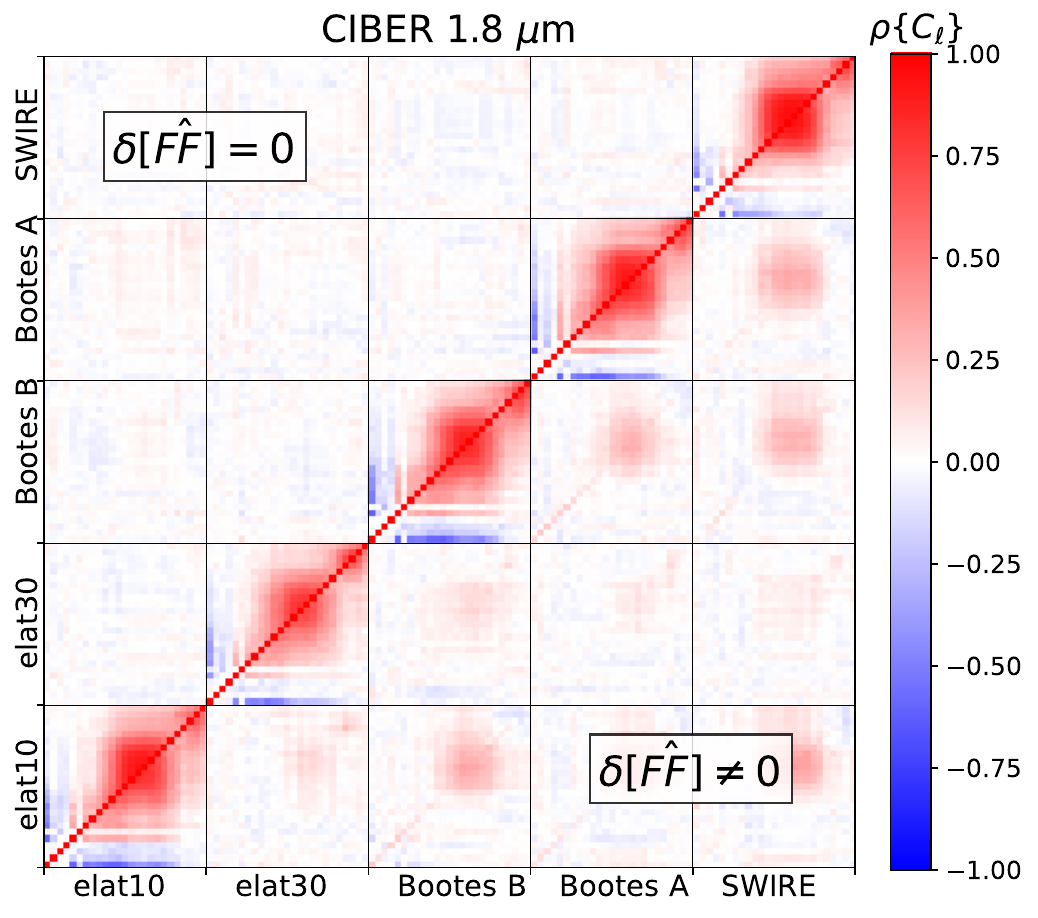}
    \caption{Mock correlation matrices for \emph{CIBER} 1.1 $\mu$m (left) and 1.8 $\mu$m (right), where $\ell$ runs from low to high in each sub-block. The upper triangular component of each matrix is derived from the case with no FF errors ($\delta[\hat{FF}]=0$), while the lower triangular component shows the same elements for the case with FF errors ($\delta[\hat{FF}]\neq 0$). The correlation matrices are derived from 1000 sets of mock \emph{CIBER} observations including instrumental noise, foregrounds, masking, filtering and, for the $\delta[\hat{FF}]\neq 0$ case, FF estimation. These observational effects induce strong mode coupling on intermediate and small scales within individual fields and between pairs of fields. The mild correlations between fields in the $\delta[\hat{FF}]=0$ case reflect deviations from each weighted field average, which contains information from all fields; these correlations approach zero in the limit of more simulations.}
    \label{fig:covariance_matrices}
\end{figure*}

For each test configuration we calculate the bandpower covariance matrix describing departures of each field's recovered power spectrum (indexed by $i$) with respect to the field average from set $j$:
\begin{equation}
    \hat{\mathcal{C}}_{mock} = \langle (C_{\ell, i}^j - C_{\ell, av}^j)^2\rangle.
\end{equation}
Note that this is different than the covariance computed relative to the true underlying sky power spectra, and is used to test data consistency in the observed data. To highlight the difference in correlation structure with and without FF errors, we show the correlation matrices $\rho(\lbrace{\hat{C}_{\ell}\rbrace})$ for both bands in Fig. \ref{fig:covariance_matrices}. In each plot, the upper triangular component is the correlation matrix for $\delta[\hat{FF}]=0$, while the lower triangular component shows the full $\delta[\hat{FF}]\neq 0$ case. The within-field bandpower covariance (block-diagonal matrices) shows similar structure in both cases, with strong correlations from mode coupling and read noise on intermediate and small scales. These correlations are stronger for 1.1 $\mu$m than 1.8 $\mu$m which we attribute to the different noise levels across the two bands. Unlike the $\delta[\hat{FF}]=0$ case, for which each field is treated separately, the $\delta[\hat{FF}]\neq 0$ case shows significant correlations between fields. The cross-field covariance arises because of our FF correction, which mixes the information from all the fields into each field's power spectrum estimate. This demonstrates the importance of accounting for the full field-field covariance when assessing internal consistency of observed auto-power spectra.

Within each field, we observe a strong anti-correlation between the lowest $\ell$ bandpower and intermediate scale bandpowers. We determine this to be the result of fitting per-quadrant offsets to mitigate detector effects, as we do not observe the anti-correlation in tests without the per-quadrant offset fitting. 

\subsection{Impact of filtering on recovered CIBER power spectra}

We assess the impact of image filtering methods on recovered CIBER power spectra. As seen in Fig. \ref{fig:t_ell}, separating a best-fit, \nth{2}-order Fourier component model from the maps nearly nulls the lowest two bandpowers, with milder suppression in higher $\ell$ bandpowers. When incorporated into the mode mixing matrices, we find that the second column of the matrix is of order $10^{-18}$, which makes inversion of the full mixing matrix unstable. We address this by truncating each matrix, excluding the rows and columns involving the lowest two bandpowers, and applying the inverse of the truncated matrix to the upper twenty-three measured bandpowers. 

\begin{figure*}
    \centering
    \includegraphics[width=0.49\linewidth]{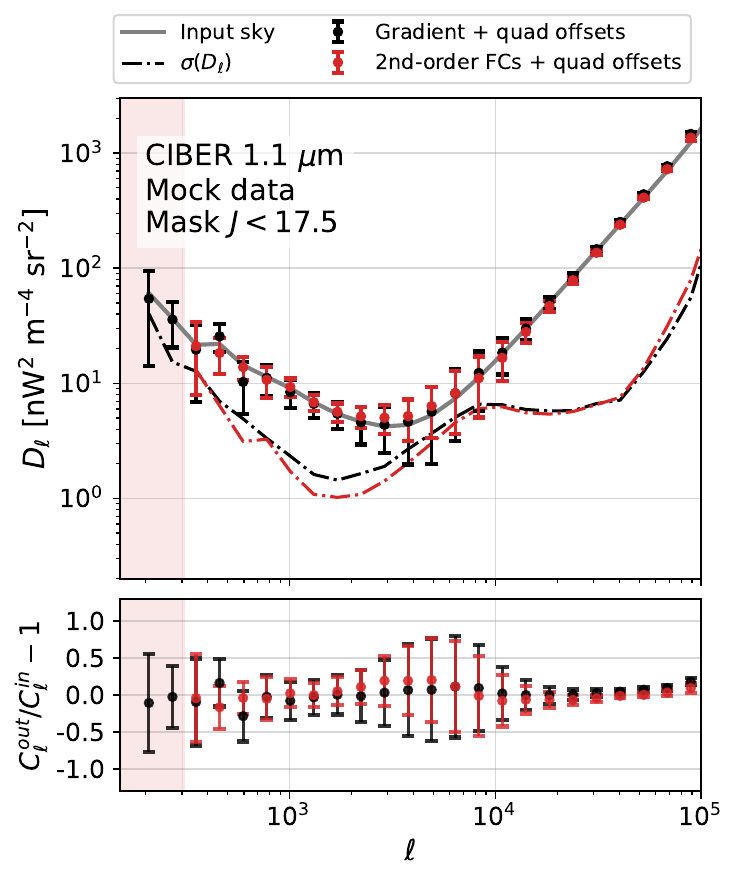}
    \includegraphics[width=0.49\linewidth]{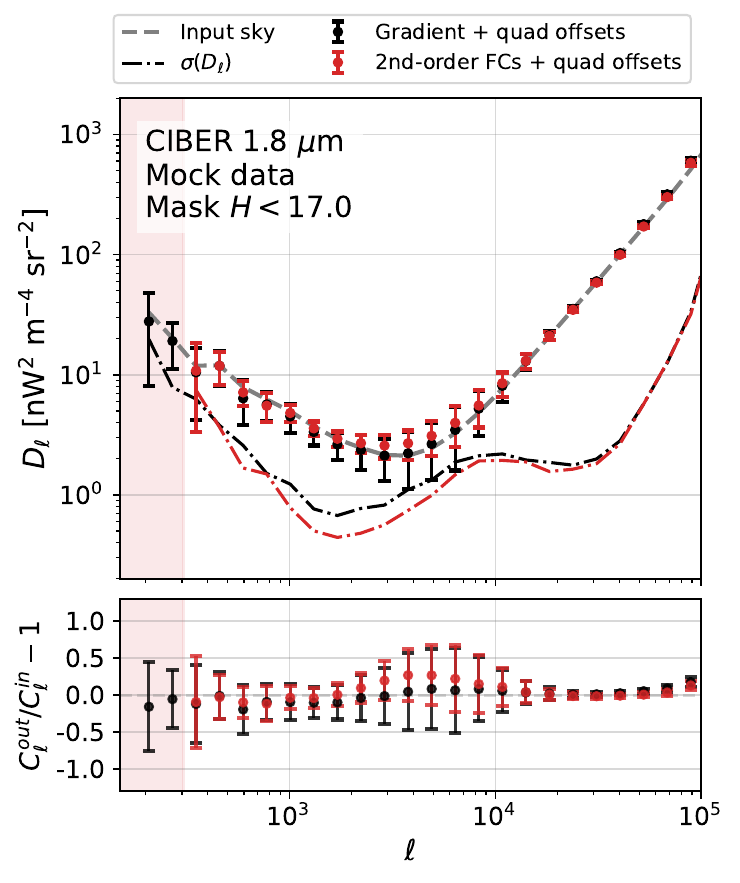}
    \caption{Field-averaged power spectrum recovery for two kinds of image filtering: per-quadrant offsets + gradient (black), and per-quadrant offsets + \nth{2}-order Fourier component model (red). The latter is a more aggressive filter on large angular scales, effectively nulling the lowest two bandpowers (indicated by the shaded red regions). We show 1$\sigma$ power spectrum uncertainties for each case, which we calculate as the dispersion of recovered power spectra from 1000 mocks (dash-dotted lines).}
    \label{fig:mock_cl_vs_filter}
\end{figure*}

In Figure \ref{fig:mock_cl_vs_filter} we compare our fiducial results and 1$\sigma$ uncertainties (black) with those using the Fourier component filtering (red). On scales $2500 < \ell < 10000$, the recovered power spectra using Fourier component filtering exhibit a positive bias with a level of $10-25\%$. These bandpowers are expected to be highly correlated (c.f. Fig. \ref{fig:covariance_matrices}). We attribute this bias to our mixing matrix truncation \textemdash\ while the Fourier component filtering reduces the off-diagonal mixing matrix components of the first bandpower by a factor of several orders of magnitude relative to gradient filtering, there is a small amount of residual leakage of low-$\ell$ power to smaller scale modes. On larger angular scales, our Fourier component filtering approach recovers unbiased fluctuations that are comparable to those using gradient filtering (and more unbiased for a handful of bandpowers).

While the \nth{2}-order Fourier component filter effectively removes information from the lowest two bandpowers, we find that the corresponding dispersion of recovered power spectra in the range $500 < \ell < 2000$ is $30-40$\% smaller than that using gradient filtering, with more improvement for 1.8 $\mu$m than 1.1 $\mu$m. We find that the improved sensitivity on several arcminute scales is due to suppression of low-$\ell$ sample variance, which otherwise propagates to smaller scales. The CIBER measurements from \zem\ suggest the presence of sky fluctuations with a red spatial index on scales $\ell \lesssim 1000$, in which case this sample variance effect is particularly relevant. We confirm this hypothesis by testing both filtering cases on simulations with instrument noise and IGL+ISL but no $\ell^{-3}$ fluctuation component. In this test configuration, we recover unbiased power spectra with no change in sensitivity.

\begin{figure}
    \centering
    \includegraphics[width=0.48\linewidth]{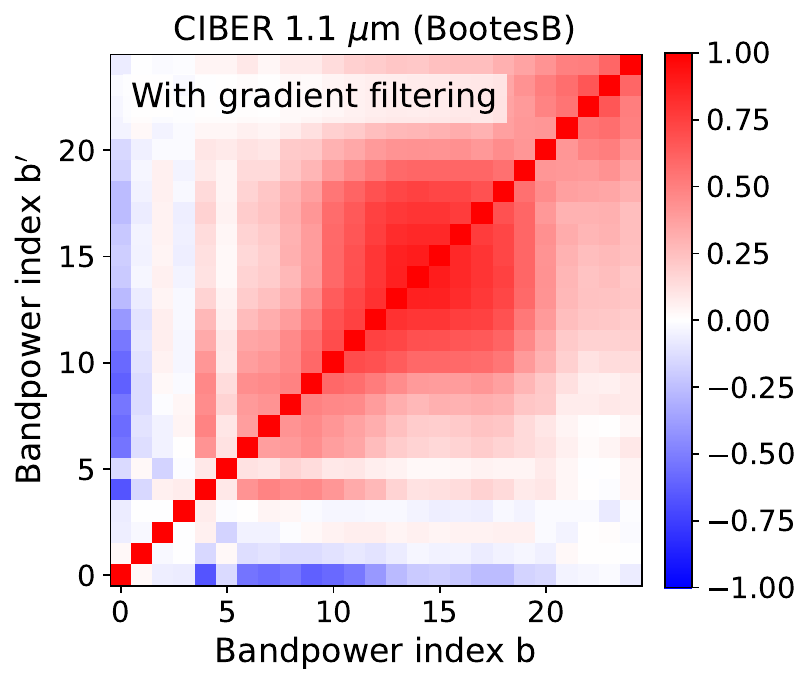} \includegraphics[width=0.48\linewidth]{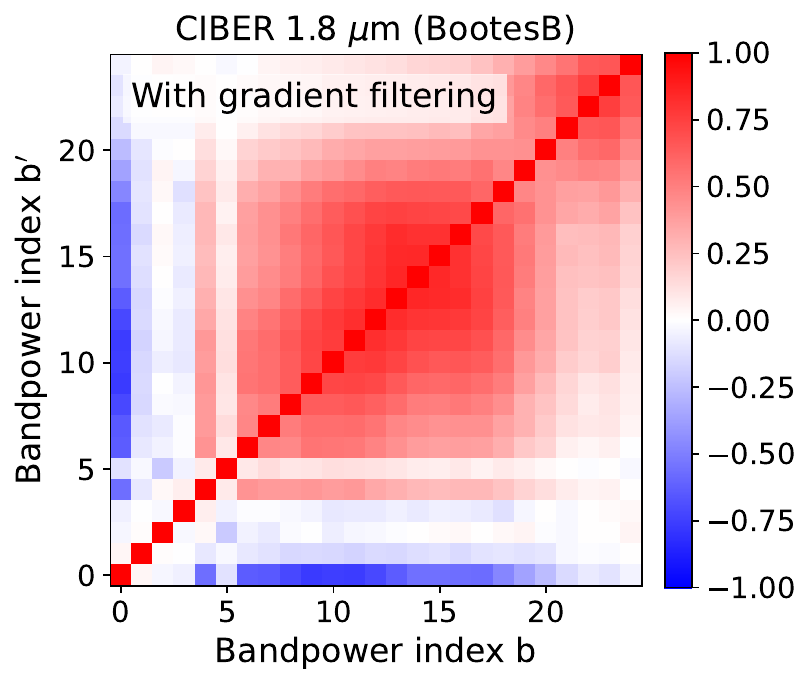} 
    \includegraphics[width=0.48\linewidth]{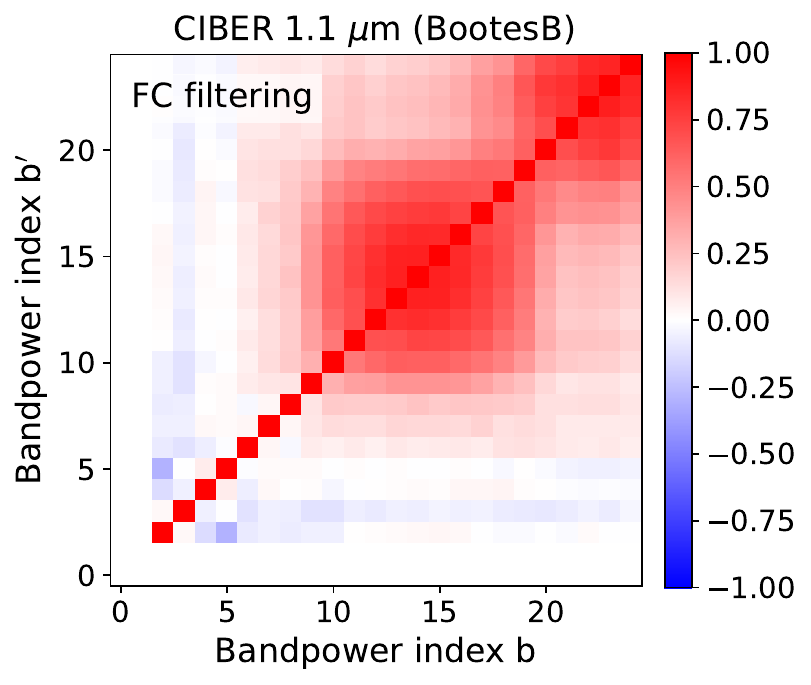} 
    \includegraphics[width=0.48\linewidth]{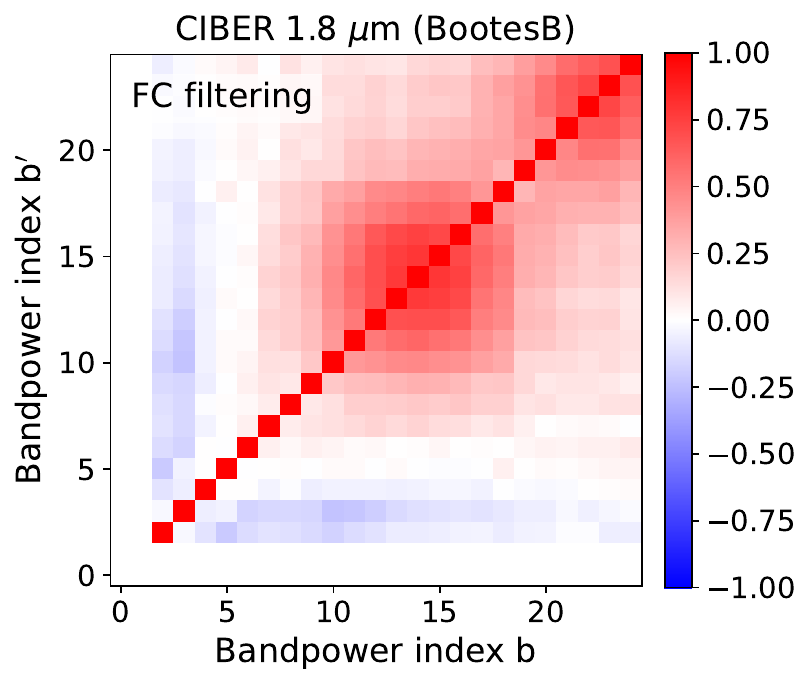} 
    \caption{Comparison of bandpower correlation coefficient matrices $\rho(C_{\ell})$ for the Bootes B field for 1.1 $\mu$m (left column) and 1.8 $\mu$m (right), each derived from an ensemble of 1000 recovered mock power spectra.}
    \label{fig:corr_matrix_vs_filter}
\end{figure}

In Figure \ref{fig:corr_matrix_vs_filter} we compare the correlation matrices of recovered power spectra for one of our fields (Bootes B) using our two filtering configurations. For both bands we find that Fourier component filtering reduces the presence of sharp correlations across modes as well as the overall correlation coefficients. 

\subsection{Power spectrum recovery for varying masking depths}
\label{sec:psmock_vs_maskmag}

Having demonstrated power spectrum estimation for our fiducial masking case, we now test our full pipeline on the same mocks but over a much broader range of source masking thresholds. This includes recovery of both point source-dominated power spectra (i.e., shallow masking cuts) and much deeper cuts, in total spanning three orders of magnitude in Poisson fluctuation power. For sources with $J<11$, non-linear detector response and saturation in the observed data preclude reliable measurements without detailed corrections. Our deepest masking cuts ($J<18.5$ and $H<18.0$ for 1.1 $\mu$m and 1.8 $\mu$m, respectively) are informed by the reliability of our source masking algorithm as demonstrated in \S \ref{sec:mask}. Although it is not our science focus to measure point source-dominated Poisson fluctuations, this exercise enables us to test the consistency of large-angle fluctuations in the observed data as a function of masking depth.

We note that our matrix formalism breaks down in the presence of bright unmasked point sources. This is a result of using the FF stacking estimator, in which bright point sources need to be masked regardless of masking depth to avoid large FF errors. In practice we use the $J<17.5$ and $H<17.0$ masks to calculate $\hat{FF}$ for all shallower masking cuts). As a result, the FF errors driven by sky signal differ from that of the target signal (which does contain bright point sources), meaning that the linear FF bias correction is not exact. In place of a full treatment, which would require an iterative or simultaneous estimation of power spectra at several masking depths, we characterize this effect empirically using the mocks. We determine that the $M_{\ell\ell^{\prime}}$ correction without FF errors recovers more accurate power spectra down to $(J_{lim},H_{lim})=15$, while $M_{\ell\ell'}^{mask+FF+filt}$ is more accurate for deeper source masking cuts.

\begin{figure*}
\centering
\includegraphics[width=0.49\linewidth]{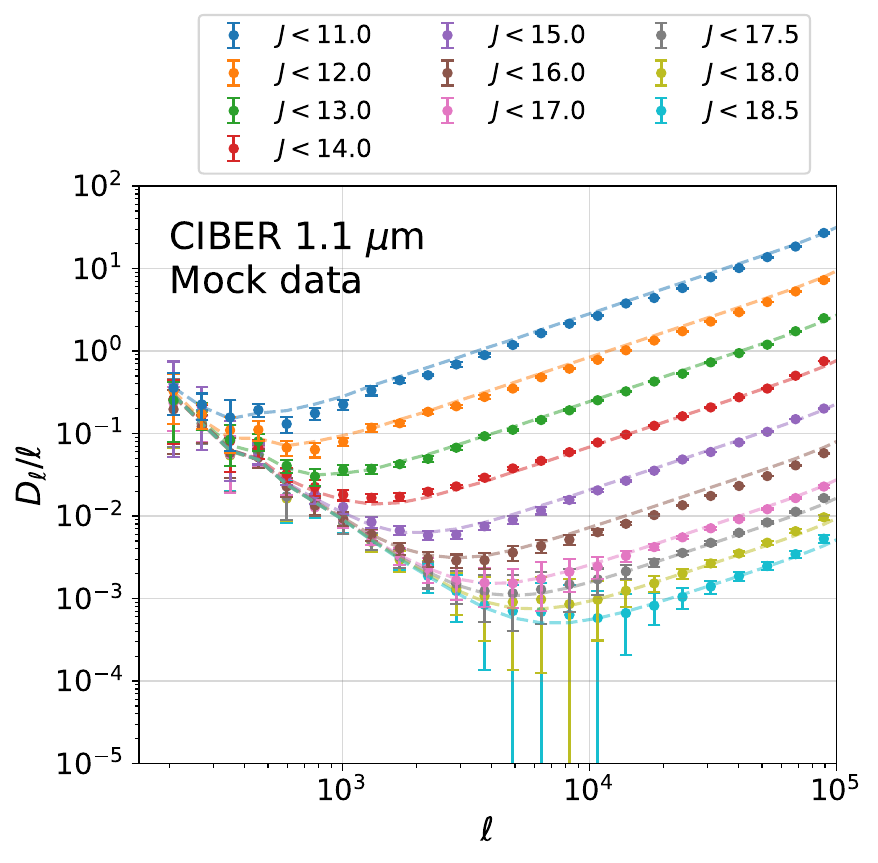}
\includegraphics[width=0.49\linewidth]{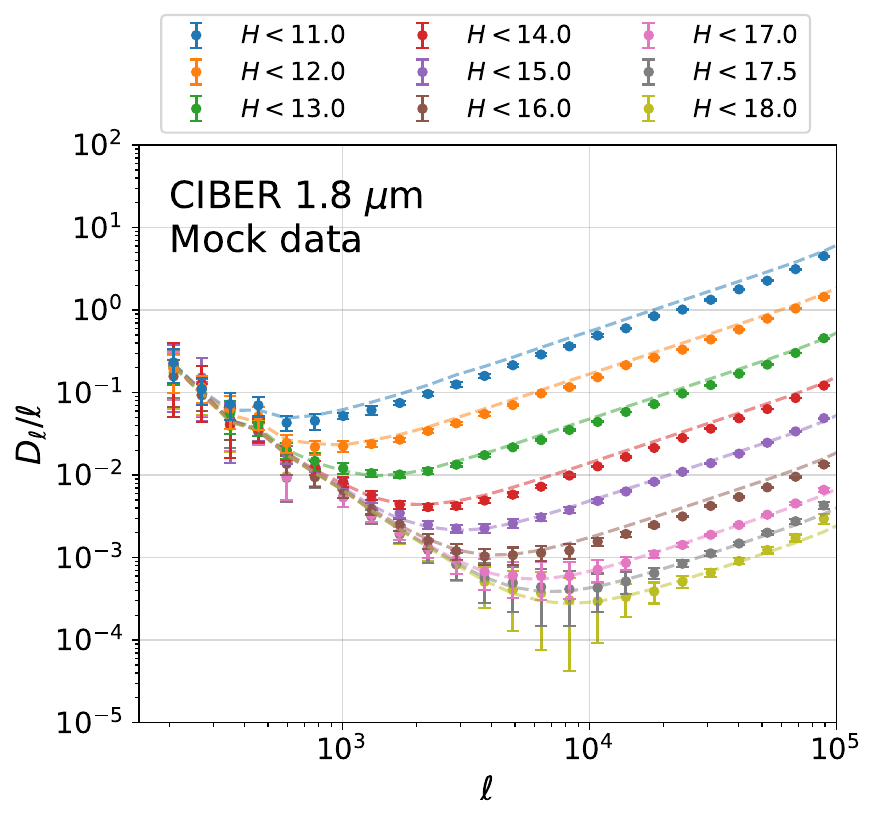}
\caption{Mock power spectrum recovery for a range of masking cuts using the simulations described in \S \ref{Sec:mocks}. These results validate our ability to recover unbiased power spectra in the presence of Poisson noise spanning three orders of magnitude in power.}
\label{fig:mock_ps_recovery_vs_mag}
\end{figure*}

We show the recovered power spectra as a function of masking depth in Figure \ref{fig:mock_ps_recovery_vs_mag}. These results validate our ability to measure large-angle fluctuations across all masking cases. As expected, the fractional power spectrum errors on scales $\ell > 1000$ are largest near the masking depth where we transition from $M_{\ell\ell'}^{mask+filt}$ to $M_{\ell\ell'}^{mask+filt+FF}$. The slight underestimation for $(J_{lim},H_{lim})=16$ is due to the fact that the masks used to estimate the FF are deeper than those used to compute the power spectrum, such that $M_{\ell\ell'}^{mask+filt+FF}$ slightly overcorrects the target signal. We do not pursue shot noise corrections using the estimated power spectra at the FF masking depth, however this may be important in settings with stronger requirements on estimation of Poisson fluctuations.

\section{Conclusion}
\label{sec:conclusion}

In this work we present an extension of the pseudo-$C_{\ell}$ formalism for measurements of NIR EBL anisotropies, with application to imaging data from the Cosmic Infrared Background Experiment. Improving on the methodology in \zem, we address two important effects necessary for measurement of sky fluctuations, namely deep source masking and in flight FF correction. We derive sky flats directly from the science fields and build on the pseudo-$C_{\ell}$ formalism to correct for additive and multiplicative biases sourced by FF errors. Through tests on mock \emph{CIBER} observations with injected FFs obtained from laboratory measurements, we demonstrate that our power spectrum pipeline can recover unbiased power spectra for all but the smallest angular scales. Because we only have five fields, the flat-fielding process increases our power spectrum uncertainties. By comparing against similar mock tests where the FF is assumed to be known perfectly, we determine that residual FF errors increase uncertainties on scales $500<\ell<2000$ by less than 20\%. Our scheme bypasses the use of field differences used in \zem, which opens the potential for more aggressive point source masking in individual fields. Our source masking approach is an efficient, data-driven alternative to direct spectral energy distribution fitting, and enables us to mask two magnitudes deeper in the NIR than is possible with existing 2MASS photometry in the \emph{CIBER} fields. These improvements lead to new measurements of clustering fluctuations on several arcminute to degree scales in both auto and cross-power spectra at 1.1 $\mu$m and 1.8 $\mu$m, which we present in a companion paper.

Correcting for mode coupling effects is an important component of this work. As NIR EBL fluctuation measurements become signal dominated there are a number of additional pseudo-$C_{\ell}$ corrections that will be important to consider. The first involves the fact that many filtering operations can couple with the astronomical masks, and should be folded into $M_{\ell\ell'}$ estimation. In this work we incorporate image filtering into our mode mixing corrections. The second effect, which we do not correct for, involves the dependence of the mode coupling correction on the shape of the underlying sky power spectrum. By choosing sufficiently fine bandpowers one can mitigate biases from this effect, however in general the mode coupling will be affected by derivatives of the sky power spectrum $\lbrace\delta^i C_{\ell}/\delta \ell^i\rbrace$. The shape of the power spectrum can be incorporated as a prior in mode coupling corrections through the bandpower operator $P_{bl}$, as done in \cite{master_leung}. It has been shown that the standard MASTER result for computing pseudo-$C_{\ell}$ estimates is biased when there is correlation between the signal and the mask \citep{cheng22, cmb_maskcorr}. \cite{remaster} shows this correlation can be calculated analytically and through simulations, with corrections going as the three- and four-point functions of the maps and masks. Estimating such corrections requires the use of more realistic sky models, such as those from MICECAT, since they may be sensitive to the one-halo contributions of IGL and any additional intra-halo light. 

In some cases, mode coupling effects can be mitigated at the map level.
For example, the fraction of masked pixels can be reduced by subtracting bright stars (or extended PSF components), given accurate knowledge of the PSF and source position. With redshift information, which will be available for many SPHEREx galaxies \citep{feder23}, it may be possible to perform physically motivated masking, in which the extended light component of bright, low-redshift galaxies can be removed at fixed comoving radius. Targeted point source/CIB de-projection \citep{act_deproj, mccarthy_deproj} or more general component separation of pointlike and diffuse signals \citep{pcatde} are other avenues toward mitigating the effects of mode coupling on future fluctuation measurements. 

The significance of these effects (and mitigation techniques) can be tested directly through tests on realistic mocks in order to assess their impact at any given experiment sensitivity. While we treat power spectrum estimation as an inverse problem, it may be more reliable in future NIR EBL inferences to instead forward model pseudo-$C_{\ell}$ measurements, returning $C_{\ell}$ \emph{reconstructions} rather than inverted $C_{\ell}$ point estimates. This requires careful consideration of instrumental and observational effects, but also allows the incorporation of more realistic EBL simulations with proper correlations in the mock intensity maps. Such a Bayesian approach offers a promising path to obtaining interpretable fluctuation measurements.

Existing and near-future experiments will map out the NIR EBL over larger regions of sky with significantly broader spectral coverage and resolving power. \emph{CIBER}-2, the second generation of \emph{CIBER}, has three H2RG detectors and six windowpane filters for imaging at $0.5-2.5$ $\mu$m \citep{nguyen_16, shirahata_16, takimoto_ciber2}. The power spectrum formalism from this work will be important for \emph{CIBER}-2 data, which is similar in structure to that of \emph{CIBER}-1. The Spectro-Photometer for the History of the Universe, Epoch of Reionization, and Ices Explorer (SPHEREx) will conduct a two-year, all-sky survey in 102 bands spanning $0.75-5$ $\mu$m, dramatically increasing the sensitivity and volume of NIR broad-band intensity mapping data. The primary focus for SPHEREx will be the $\sim$200 deg$^2$ centered near the ecliptic poles. A daily cadence over the poles throughout the two-year survey will enable accurate in-flight estimates of the instrument FF response and dark current. Diffuse light measurements will also be pursued with imager data from the \emph{Euclid} mission through the LIBRAE project \citep{librae}. The pseudo-$C_{\ell}$ formalism and simulation-based approach presented in this work can be built to characterize residual systematic uncertainties and more complicated observational phenomena through forward models that match the requisite realism of the data generating process.  

Improved methods to estimate CNIB fluctuations, a larger analysis toolkit to interpret measurements and a dramatic increase in data quantity and quality will transform our ability to study the history of cosmic light production in the coming years, uncovering features of the low surface brightness universe that may yield unanticipated insights about galaxy evolution and large-scale structure formation.

\acknowledgements

We thank Bryan Steinbach for helpful discussions regarding all things mode coupling-related.

This material is based upon work supported by the National Aeronautics and Space Administration under APRA research grants NNX10AE12G, NNX16AJ69G, 80NSSC20K0595, 80NSSC22K0355, and 80NSSC22K1512. 

This publication makes use of data products from the Two Micron All Sky Survey, which is a joint project of the University of Massachusetts and the Infrared Processing and Analysis Center/California Institute of Technology, funded by the National Aeronautics and Space Administration and the National Science Foundation.

The Pan-STARRS1 Surveys (PS1) and the PS1 public science archive have been made possible through contributions by the Institute for Astronomy, the University of Hawaii, the Pan-STARRS Project Office, the Max-Planck Society and its participating institutes, the Max Planck Institute for Astronomy, Heidelberg and the Max Planck Institute for Extraterrestrial Physics, Garching, The Johns Hopkins University, Durham University, the University of Edinburgh, the Queen's University Belfast, the Harvard-Smithsonian Center for Astrophysics, the Las Cumbres Observatory Global Telescope Network Incorporated, the National Central University of Taiwan, the Space Telescope Science Institute, the National Aeronautics and Space Administration under Grant No. NNX08AR22G issued through the Planetary Science Division of the NASA Science Mission Directorate, the National Science Foundation grant No. AST-1238877, the University of Maryland, Eotvos Lorand University (ELTE), the Los Alamos National Laboratory, and the Gordon and Betty Moore Foundation.

\software{\texttt{matplotlib}, \texttt{numpy}, \texttt{scipy}, \texttt{sklearn}}
\appendix
\section{Power spectrum estimation with FF stacking estimator}
\label{sec:ff_formalism}

\subsection{Stacking FF estimator}
In this Appendix we introduce the stacking FF estimator and propagate errors in the FF to power spectrum biases. To start, consider a single sky realization $I_i(x,y)$ for field $i$, the sum of a mean normalization $\overline{I}_i^{sky}$ (dominated by ZL), and a fluctuation component $S_i$ which is composed from EBL sky fluctuations, diffuse galactic light (DGL) fluctuations and integrated stellar light (ISL):
\begin{equation}
    I_i^{sky}(x,y) = \overline{I}_i^{sky} + S_i = \overline{I}_i^{sky} + S_i^{EBL} + S_i^{DGL} + S_i^{ISL}.
    \label{eq:sky_signal}
\end{equation}
The FF responsivity is defined as a scalar field $FF(x,y)$ for each detector. The incident sky signal has associated photon noise $\epsilon_{\gamma}$, and this signal+noise component is multiplied by $FF(x,y)$, after which read noise (denoted by $\epsilon_{read}$) is imprinted, producing the observed image $I_i^{obs}(x,y)$: 
\begin{equation}
    I_i^{obs}(x,y) = FF(x,y)\left[\overline{I}_i^{sky} + S_i^{EBL} + S_i^{DGL} + S_i^{ISL}+ \epsilon_{\gamma,i}\right] + \epsilon_{read,i}.
\end{equation}

The FF estimate derived from field $i$ is obtained by dividing the observed image by the mean surface brightness in unmasked pixels:
\begin{align}
    \hat{FF}_i(x,y) &= \frac{I_i^{obs}(x,y)}{\overline{I}_i^{sky}} = \frac{FF(x,y)\left[\overline{I}_i^{sky} + S_i^{EBL} + S_i^{DGL} + S_i^{ISL} + \epsilon_{\gamma,i}\right] + \epsilon_{read,i}}{\overline{I}_i^{sky}}.
\end{align}
The error on the FF estimate for a single field can be expressed in terms of the assumed sky and noise components:
\begin{equation}
    \delta \hat{FF}_i(x,y) = \frac{FF(x,y)(S_i^{EBL}+S_i^{DGL} + S_i^{ISL}+ \epsilon_{\gamma,i}) + \epsilon_{read,i}}{\overline{I}_i^{sky}}.
\end{equation}
This highlights that non-uniform fluctuation components lead to errors in $\hat{FF}$. The FF responsivity only needs to be evaluated in unmasked pixels,  but each pixel has a specific number of off-field measurements, which depends on the masks in the off field stacks,
\begin{equation}
    \delta \hat{FF}_i(x,y) = M_i(x,y)\left(\frac{FF(x,y)(S_i^{EBL}+S_i^{DGL} + S_i^{ISL}+ \epsilon_{\gamma,i}) + \epsilon_{read,i}}{\overline{I}_i^{sky}}\right).
\end{equation}

We apply pixel weights unique to each off-field used in a given stacked FF estimate. Denote the per-pixel RMS of the FF error by $\sigma$. The resulting inverse variance weights are

\begin{equation}
w_i \approx \left(\frac{\overline{I}_i^{sky}}{\sigma_{tot}}\right)^2,
\end{equation}
where
\begin{equation}
    \sigma_{tot} = \sqrt{\sigma^2_{\gamma} + \sigma_{EBL}^{2}+\sigma_{DGL}^{2} + \sigma_{read}^2}.
\end{equation}
Assuming the per-pixel fluctuations from the sky signal $S_i$ are subdominant to the instrumental noise, we can approximate the weights as
\begin{equation}
    w_i \approx \left(\frac{\overline{I}_i^{sky}}{\sqrt{\sigma^2_{\gamma,i} + \sigma^2_{read,i}}} \right)^2.
\end{equation}
Assuming our field weights are sum-normalized, i.e., $\sum_i w_i = 1$, the stacked FF $\hat{FF}^j(x,y)$ is
\begin{equation}
    \hat{FF}^j(x,y) = \sum_{i=1}^{N_f}w_i \hat{FF}_i.
\end{equation}
The $N_f=4$ fields that go into each stacked \emph{CIBER} FF estimate are uncorrelated, such that the variance of the stacked FF can be written as a weighted sum of variances
\begin{equation}
    \textrm{Var}[\hat{FF}^j] = \sum_{i=1}^{N_f}w_i^2 \textrm{Var}[\hat{FF}_i].
\end{equation}
Condensing the sky fluctuation signal for field $i$ into $S_i$ and using inverse variance weights, we write the FF standard error as
\begin{equation}
    \delta[\hat{FF}]_j = \left(\sum_{i\neq j}\delta[\hat{FF}]_i^{-2}\right)^{-1/2}
    = \left(\sum_{i\neq j}\left[\frac{M_i\left(FF(S_i+\epsilon_{\gamma,i})+\epsilon_{read,i}\right)}{\overline{I}_i^{sky}} \right]^{-2} \right)^{-1/2}.
\end{equation}
This stacking estimator allows us to model FF errors with minimal assumptions on the underlying fluctuations. This is opposed to using lab-derived FFs, which assume a perfectly uniform illuminating surface but have errors that are harder to quantify.

\subsection{Flat field bias}

We now quantify how FF errors propagate to estimates of the auto-power spectrum. After deriving the generic expression for each FF-corrected image, we separate FF error contributions driven by instrument noise (\ref{sec:app_ff_noise_bias}) and sky fluctuations (\ref{sec:unmasked_ffbias}). In the absence of noise (or after proper noise bias subtraction), FF errors from sky fluctuations lead to a multiplicative bias on the recovered power spectrum, assuming the fluctuations from multiple fields are drawn from the same underlying distribution. We start with the unmasked case and detail the role and treatment of mode coupling in \ref{sec:ffmask}.

For this calculation we assume that foreground point sources are perfectly removed from the maps. Following Eq. \eqref{eq:sky_signal} the observed signal for field $j$ is
\begin{equation}
    I_j^{obs} = FF_{true}\left[I_j^{sky} + \epsilon_{\gamma,j} \right] + \epsilon_{read,j}
\end{equation}
Let us express the FF estimate in terms of the true FF and the FF error, i.e.,
\begin{equation}
    \hat{FF}^j = FF_{true} + \delta [\hat{FF}^j] = FF_{true}\left[1 + \frac{\delta [\hat{FF}^j]}{FF_{true}}\right]. 
\end{equation}
Then the FF-corrected image can be written as
\begin{equation}
    \frac{I_j^{obs}}{\hat{FF}^j} = \frac{FF_{true}\left[I_j^{sky}+\epsilon_{\gamma,j}\right] + \epsilon_{read,j}}{FF_{true}\left[1 + \frac{\delta [\hat{FF}^j]}{FF_{true}}\right]}.
    \label{eq:iobs_ffhat}
\end{equation}
Assuming $\delta [\hat{FF}^j]/FF_{true} \ll 1$, we Taylor expand \eqref{eq:iobs_ffhat}:
\begin{align}
    \frac{I_j^{obs}}{\hat{FF}^j} &\approx \left(I_j^{sky}+\epsilon_{\gamma,j} + \frac{\epsilon_{read,j}}{FF_{true}}\right)\left(1 - \frac{\delta [\hat{FF}^j]}{FF_{true}}\right) \\
    &= I_j^{sky} + \epsilon_{\gamma,j} +\frac{\epsilon_{read,j}}{FF_{true}} - \frac{\delta [\hat{FF}^j]}{FF_{true}}\left(I_j^{sky} + \epsilon_{\gamma,j} + \frac{\epsilon_{read,j}}{FF_{true}}\right).
    \label{eq:taylor_expand_ffcorr}
\end{align}
We now rearrange terms, expressing $FF_{true}$ in terms of the $\hat{FF}$ and $\delta[\hat{FF}]$. For example,
\begin{equation}
    \frac{\epsilon_{read,j}}{FF_{true}} = \frac{\epsilon_{read,j}}{\hat{FF}^j - \delta[\hat{FF}^j]} = \frac{\epsilon_{read,j}}{\hat{FF}^j(1 - \frac{\delta[\hat{FF}^j]}{\hat{FF}^j})} \approx \frac{\epsilon_{read,j}}{\hat{FF}^j}\left(1 + \frac{\delta[\hat{FF}^j]}{\hat{FF}^j}\right)
\end{equation}
such that

\begin{align}
    \frac{I_j^{obs}}{\hat{FF}^j} &\approx I_j^{sky}+\epsilon_{\gamma,j} + \frac{\epsilon_{read, j}}{\hat{FF}^j}\left(1+\frac{\delta[\hat{FF}^j]}{\hat{FF}^j}\right) - \frac{\delta[\hat{FF}^j]}{\hat{FF}}\left(1+\frac{\delta[\hat{FF}^j]}{\hat{FF}^j}\right)\left(I_j^{sky} + \epsilon_{\gamma,j} + \frac{\epsilon_{read,j}}{FF_{true}}\right) \\
    &= I_j^{sky}+\epsilon_{\gamma,j} + \frac{\epsilon_{read, j}}{\hat{FF}^j} + \frac{\epsilon_{read,j}}{\hat{FF}^j}\frac{\delta[\hat{FF}^j]}{\hat{FF}} - \frac{\delta[\hat{FF}^j]}{\hat{FF}^j}\left(I_j^{sky}+\epsilon_{\gamma,j} + \frac{\epsilon_{read,j}}{FF_{true}}\right) + \mathcal{O}(\delta^2 [\hat{FF}^j]) \\
    &\approx I_j^{sky}+\epsilon_{\gamma,j} + \frac{\epsilon_{read, j}}{\hat{FF}^j} + \frac{\delta[\hat{FF}^j]}{\hat{FF}^j}\left(I_j^{sky}+\epsilon_{\gamma,j}\right) + \left[\frac{\epsilon_{read,j}}{\hat{FF}^j}\frac{\delta[\hat{FF}^j]}{\hat{FF}^j} - \frac{\epsilon_{read,j}}{FF_{true}}\frac{\delta[\hat{FF}^j]}{\hat{FF}^j}\right] \\
    &\approx I_j^{sky}+\epsilon_{\gamma,j} + \frac{\epsilon_{read, j}}{\hat{FF}^j} + \frac{\delta[\hat{FF}^j]}{\hat{FF}^j}\left(I_j^{sky}+\epsilon_{\gamma,j}\right) + \mathcal{O}(\delta^2 [\hat{FF}^j]).
\end{align}
After discarding all terms of order $\delta^2[\hat{FF}^j]$, we find
\begin{align}
\label{eq:flatnoise}
    \frac{I_j^{obs}}{\hat{FF}^j} &\approx \overline{I}_j^{sky} + S_j + \epsilon_{\gamma,j} + \frac{\epsilon_{read,j}}{\hat{FF}^j} - \frac{\delta[\hat{FF}^j]}{\hat{FF}^j}(\overline{I}_j^{sky}+S_j+\epsilon_{\gamma,j}).
\end{align}
This expression highlights contributions at the map level sourced by the noisy FF correction that modify the measured pseudo-power spectrum. The fractional FF error $\delta[\hat{FF}^j]/\hat{FF}^j$ coupled to the mean sky brightness $\overline{I}_j^{sky}$ is the leading additional contribution to the pseudo-power spectrum.

\subsubsection{Noise bias subtraction}
\label{sec:app_ff_noise_bias}
We assume that FF errors from read noise and photon noise in each stack are uncorrelated with the signal, such that the corresponding noise power adds linearly. We define the FF estimate in the absence of sky fluctuations as $\hat{FF}_{inst}$:
\begin{equation}
    \hat{FF}^j_{inst} = FF\left[1 + \frac{\epsilon_{\gamma,j}}{\overline{I}_j^{sky}}\right] + \frac{\epsilon_{read,j}}{\overline{I}_j^{sky}}.
\end{equation} 
After correcting the mean sky brightness by $\hat{FF}_{inst}$,
\begin{equation}
    \frac{\overline{I}_j^{sky}}{\hat{FF}_{inst}^j} = \overline{I}_j^{sky} + \epsilon_{\gamma,j} + \frac{\epsilon_{read,j}}{\hat{FF}^j_{inst}} - \frac{\delta[\hat{FF}_{inst}^j]}{\hat{FF}^j_{inst}}\left(\overline{I}_j^{sky} + \epsilon_{\gamma,j}\right).
    \label{eq:nofluc_noise}
\end{equation}

We can express the FF error $\delta[\hat{FF}^j]$ in terms of its instrument noise and sky fluctuation components, 
\begin{equation}
    \delta[\hat{FF}^j] = \delta[\hat{FF}^j_{inst}] + \delta[\hat{FF}^j_S].
\end{equation}
Using this we expand \eqref{eq:flatnoise} and subtract the noise bias terms from \eqref{eq:nofluc_noise}, discarding terms $\mathcal{O}(\delta \hat{FF}^j_{inst} \delta \hat{FF}^j_S)$.
\begin{equation}
    \frac{I_j^{obs}}{\hat{FF}^j} - \frac{\overline{I}_j^{sky}}{\hat{FF}_{inst}^j} \approx S_j - \frac{\delta[\hat{FF}^j_S]}{\hat{FF}^j}\overline{I}_j^{sky} - \left[\frac{\delta[\hat{FF}^j_S]}{\hat{FF}^j}\left(S_j + \epsilon_{\gamma,j} - \frac{\epsilon_{read,j}}{\hat{FF}^j}\right) + \frac{\delta[\hat{FF}^j_{inst}]}{\hat{FF}^j}S_j\right].
    \label{eq:noise_debiased_Iobs}
\end{equation}
In practice we compute Monte Carlo estimates of \eqref{eq:nofluc_noise} to subtract the noise bias from each observed pseudo-$C_{\ell}$ estimate. The additional noise terms in brackets are sub-dominant to the second term in \eqref{eq:noise_debiased_Iobs} and depend on the underlying fluctuations $S_j$. To assess the importance of these cross terms we evaluate them directly using sky signal mocks and noise model realizations. We confirm that they are small relative to the underlying sky power ($\lesssim 1\%$ of $C_{\ell}^{true}$) and so we do not attempt to model them in great detail when applying noise de-biasing.

\subsubsection{FF multiplicative bias}
\label{sec:unmasked_ffbias}
From \eqref{eq:noise_debiased_Iobs} it is clear that, after noise de-biasing, there is still a leading FF error contribution from $\delta[\hat{FF}_S^j]\overline{I}_j^{sky}/\hat{FF}^j$. While we do not know the sky fluctuations $S$ \emph{a priori}, the amplitude of sky fluctuations directly affects the level of FF errors using our stacked estimator. In other words, we can treat the power spectrum bias from sky fluctuations with a multiplicative correction. To illustrate this, we derive the noise-debiased power spectrum for a set of $N_f$ fields with mean surface brightnesses $\lbrace I_i\rbrace$ and noise levels $\lbrace \epsilon_i\rbrace$. For simplicity we neglect contributions from correlated read noise.

For sky fluctuations $\delta S$, we write the per-pixel FF error for each field as
\begin{equation}
    \delta FF_i = \frac{\sqrt{\epsilon_i^2 + \delta S^2}}{I_i}.
\end{equation}
The weighted variance from stacking several fields is then \begin{equation}
    \delta[\hat{FF}^j]^2 = \sum_i w_i^2\delta FF_i^2 = \sum_i w_i^2\frac{(\epsilon_i^2 + \delta S^2)}{I_i^2}
\end{equation}
Following Eq. \ref{eq:taylor_expand_ffcorr} we can express the FF corrected image for field $j$ as
\begin{align}
    I_j^{obs}/\hat{FF}^j &\approx \left(I_j + \sqrt{\epsilon_j^2 +\delta S^2}\right)\left(1 - \sqrt{\sum_i \frac{w_i^2(\epsilon_i^2 + \delta S^2)}{I_i^2}}\right) \\
    &= I_j - I_j\sqrt{\sum_i \frac{w_i^2(\epsilon_i^2 + \delta S^2)}{I_i^2}} + \sqrt{\epsilon_j^2+\delta S^2} + \mathcal{O}(I_i^{-1})
\end{align}
In power units, 
\begin{equation}
    C_{\ell,j} \approx N_{\ell,j} + \delta S^2 + \sum_i \left(\frac{w_i I_j}{I_i}\right)^2(\epsilon_i^2+ \delta S^2).
    \label{eq:bias_diagonal}
\end{equation}
where $N_{\ell,j}$ is the standard noise bias without FF errors. To calculate the full noise bias we expand $\delta [\hat{FF}_{inst}^j]$ as in Eq. \ref{eq:nofluc_noise}
\begin{equation}
    \delta [\hat{FF}_{inst}^j] = \sqrt{\sum_i \left( \frac{w_i \epsilon_i }{I_i}\right)^2}; \quad
    I_j^{inst} = I_j + \epsilon_j + \sqrt{\sum_i \left(\frac{w_i\epsilon_i I_j}{I_i}\right)^2}.
\end{equation}
Assuming the instrument noise between fields is independent, we calculate the full noise bias:
\begin{equation}
    \tilde{N}_{\ell,j} = N_{\ell,j} + \sum_i \left(\frac{w_i\epsilon_i I_j}{I_i}\right)^2
\end{equation}
After subtracting the noise bias we are left with:
\begin{equation}
    C_{\ell,j} - \tilde{N}_{\ell,j} = \delta S^2\left[1+\sum_i \left(\frac{w_i I_j}{I_i}\right)^2\right].
\end{equation}
For equal weights, noise, and ZL levels, the bias reduces to $1+\frac{1}{N_f}$. This derivation tells us that to leading order, the FF bias depends on the relative mean surface brightness of each target field compared to those used in its FF estimate. The assumption of common underlying sky fluctuations is broken by foregrounds such as ISL and DGL. However, the impact of these departures is small at current sensitivity, which we validate through tests on mocks.

\subsubsection{Mode coupling correction in presence of FF errors}
\label{sec:ffmask}
The mask couples modes contaminated by FF errors. Denote the mask for field $j$ by $M_j$, then
\begin{align*}
    \frac{I_j^{obs}}{\hat{FF}^j}
    &\approx M_j\left[\overline{I}_j^{sky} + S_j + \epsilon_{\gamma,j} + \frac{\epsilon_{read,j}}{\hat{FF}^j} - \frac{\delta[\hat{FF}^j]}{\hat{FF}^j}(\overline{I}_j^{sky}+S_j+\epsilon_{\gamma,j})\right] \\
    &= M_j\left[\overline{I}_j^{sky} + S_j + \epsilon_{\gamma,j} + \frac{\epsilon_{read,j}}{\hat{FF}^j} - \frac{\overline{I}_j^{sky}+S_j+\epsilon_{\gamma,j}}{\hat{FF}^j}\left(\sum_{i\neq j}\left[\frac{M_i\left(FF(S_i+\epsilon_{\gamma,i})+\epsilon_{read,i}\right)}{\overline{I}_i^{sky}} \right]^{-2} \right)^{-1/2}\right].
\end{align*}
This means that the FF error $\delta[\hat{FF}^j]$ for field $j$ depends on the mode coupling of the off-field masks $\lbrace M_i \rbrace$ and their sky signals $\lbrace S_i \rbrace$. Relying on the fact that the mode coupling operations are linearly separable to good approximation, we estimate the additional mode coupling through a modification of the standard MASTER algorithm.

The procedure for computing a single Monte Carlo $M_{\ell\ell^{\prime}}$ realization (with FF errors and filtering) is as follows. For each bandpower:
\begin{enumerate}
\item Generate $N_f$ Gaussian tone realizations with power spectrum set to unity within the bandpower and zero otherwise.
\item Add the mean surface brightness levels corresponding to the fields to their respective realizations.
\item Multiply each field by its respective mask.
\item For each of the $N_f$ realizations:
\begin{itemize}
\item Construct a stacked FF estimate from the other ($N_f-1$) maps, with weights $w_i$.
\item Divide the phase realization by the estimated FF.
\item Apply map filtering on the FF-corrected realization.
\item Compute the angular power spectrum from the corrected map.
\end{itemize}
\end{enumerate}
Qualitatively, we are separating the FF error in Fourier space in order to determine the mode mixing between different bandpowers. We compute 500 Monte Carlo realizations per bandpower, which is sufficiently large that statistical errors on $\langle \hat{M}_{\ell\ell^{\prime}} \rangle$ are negligible. The implementation has been optimized using the \texttt{pyfftw} package, which allocates memory for efficient computation of discrete Fourier transforms (DFTs) and inverse DFTs. For a set of $N_f=5$ fields and twenty five bandpowers, the calculation takes $\sim$ 20 minutes in wall clock time to run on a Macbook Pro with an M1 processor and 16 GB of RAM. Given the massively parallel nature of this Monte Carlo approach, it may be possible to accelerate the computation further with Graphic Processing Units (GPUs), for which discrete FFTs are well supported within the CUDA library. We defer an investigation of CPU vs. GPU performance and utilization to future work.

\subsubsection{Field weights}
\label{sec:field_weights_app}
\begin{figure}
    \centering
\includegraphics[width=0.8\linewidth]{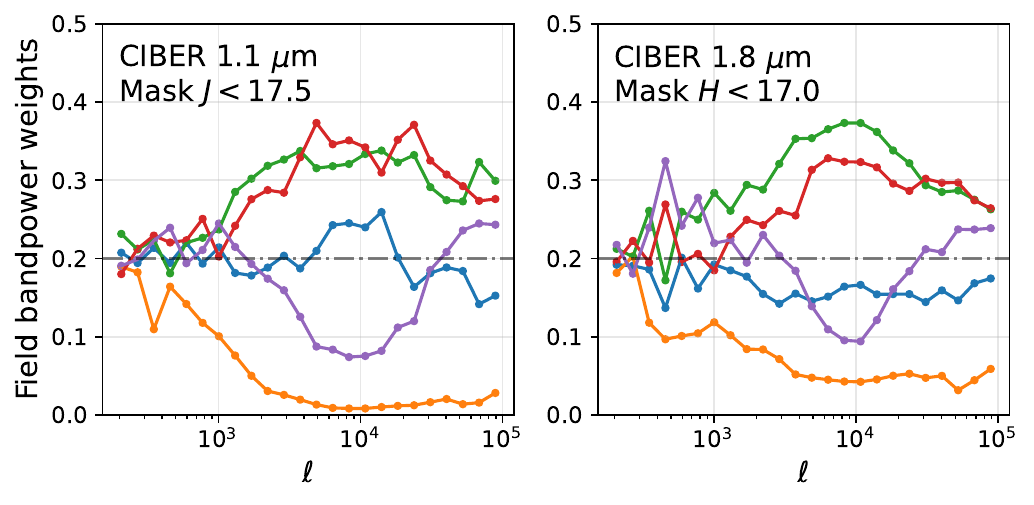}
    \caption{Inverse variance power spectrum weights for the five \emph{CIBER} fields, for 1.1 $\mu$m (left) and 1.8 $\mu$m (right). These weights are computed from the recovered power spectra of 1000 mock realizations. On intermediate scales ($5000<\ell<10000$) where read noise and mode coupling are prominent, the dispersion of the weights is largest. }
    \label{fig:field_weights_mock}
\end{figure}

The power spectrum measurement uncertainty for a given field will in general depend on the field's masking fraction, the exposure integration time and the sky brightness. We choose to combine the $C_{\ell}$ estimates across fields using optimal, per-bandpower inverse variance weights. These weights are estimated from the dispersion of recovered per-field power spectra using our ensemble of mocks. In Figure \ref{fig:field_weights_mock} we plot the derived weights as a function of multipole for our fiducial analysis. There is statistical noise at the few percent level in the weights due to the number of simulations we use. Nonetheless there is clear scale-dependence in the power spectrum weights across different fields, and the weights follow a similar scale dependence for both \emph{CIBER} bands. On large scales, our measurements are limited by statistical noise in the number of modes we sample, and so there is relatively little variation between fields. On intermediate scales where the power spectra are read noise and photon noise dominated, there is much larger variation between the field weights, with the elat30 and SWIRE fields the most downweighted. elat30 has the shortest integration time and therefore the largest instrument noise components, and so the exposure does not contain much information on these scales. SWIRE has the longest integration time and the lowest ZL level across fields, however it also has the highest source density which leads to a higher masking fraction. Masking has a large impact at intermediate scales, where noisy Fourier modes from the readout electronics mix with the masks. The relative weights for 1.1 $\mu$m and 1.8 $\mu$m share similar structure.

We use dispersion of the field weights to calculate the effective sample size from fluctuations as a function of angular scale. The fractional reduction in $N_{mode}^{eff}$ remains below 10\% on large scales ($\ell < 2000$) as well as on small scales ($\ell > 20000$). The penalty from field weights is strongest on read noise-dominated scales, where $N_{mode}^{eff}$ is 30\% and 20\% smaller for 1.1 $\mu$m and 1.8 $\mu$m respectively. 

\subsection{Cross correlation FF bias}
\label{sec:ffbias_cross}
The same FF errors that introduce a multiplicative bias in auto-power spectrum measurements also impact the \emph{CIBER} cross-power spectrum. This is due to our use of the FF stacking estimator in both \emph{CIBER} bands -- for a cross correlation between \emph{CIBER} and other instruments (e.g., from IRIS, \emph{Spitzer}) or ancillary galaxy catalogs, there is no multiplicative FF bias. However, any coherent signal across the \emph{CIBER} imagers translate to coherent FF errors across imagers, which then couple to the observed maps. As the instrument noise across imagers is uncorrelated, there is no additional noise bias from FF errors. In the unmasked case, the multiplicative bias depends on the weighted product of sky brightnesses across both bands, i.e., for bands $a$ and $b$:
\begin{equation}
    C_{\ell}^{j,\delta FF}/C_{\ell}^{j,true} = 1 + \sum_{i\neq j} \left(\frac{I_j^{a}I_j^{b}}{I_i^aI_i^b}w_i^{a}w_i^{b}\right)^2.
\end{equation}
We correct for the multiplicative bias of the cross-spectra with mode coupling using a variation on the Monte Carlo procedure detailed in \ref{sec:mode_couple_ff}.

\bibliography{references}{}
\bibliographystyle{aasjournal}

\end{document}